\documentclass[a4paper, 12pt]{article}
\pdfoutput=1

\usepackage{latexsym,amsmath,amsfonts,amssymb}
\usepackage{tikz}
\usetikzlibrary{arrows,decorations.pathmorphing,cd,decorations.markings,calc}
\usepackage{mathrsfs}
\usepackage[latin1]{inputenc}
\usepackage[american]{babel}
\usepackage{graphicx}
\usepackage{bbm}
\usepackage{cite}
\usepackage[colorlinks=true, citecolor=blue, linkcolor=blue, linktocpage=true]{hyperref}
\usepackage{tcolorbox}
\usepackage{skak}
\usepackage{enumerate}
\tikzset{snake it/.style={decorate, decoration=snake}}

\renewcommand{\baselinestretch}{1.1}
\newcommand{\Cop}{\vartheta}

\newcommand{\DEF}[2]{\vert #1 \rangle \langle #2 \vert}

\numberwithin{equation}{section}

\newcommand{\be}{\begin{equation}} \newcommand{\ee}{\end{equation}}
\newcommand{\bea}{\begin{equation} \begin{aligned}} \newcommand{\eea}{\end{aligned} \end{equation}}

\newcommand{\cA}{\mathcal{A}}

\newcommand{\cH}{\mathcal{H}}

\newcommand{\cM}{\mathcal{M}}

\newcommand{\cN}{\mathcal{N}}
\newcommand{\cO}{\mathcal{O}}

\newcommand{\cS}{\mathcal{S}}

\newcommand{\cV}{\mathcal{V}}

\newcommand{\cZ}{\mathcal{Z}}
\newcommand{\bA}{\mathbb{A}}
\newcommand{\bB}{\mathbb{B}}

\newcommand{\bZ}{\mathbb{Z}}

\newcommand{\unit}{\mathbbm{1}}

\def\repa{\raise4pt\hbox{$\square$}\mkern-14mu\raise-4pt\hbox{$\square$}}
\def\repab{\overline{\raise4pt\hbox{$\square$}\mkern-14mu\raise-4pt\hbox{$\square$}\mkern-1mu}}

\DeclareMathOperator{\sign}{sign}

\DeclareMathOperator{\abA}{\mathbb{A}}

\DeclareMathOperator{\calD}{\mathscr{D}}

\makeatletter
\DeclareFontFamily{OMX}{MnSymbolE}{}
\DeclareSymbolFont{MnLargeSymbols}{OMX}{MnSymbolE}{m}{n}
\SetSymbolFont{MnLargeSymbols}{bold}{OMX}{MnSymbolE}{b}{n}
\DeclareFontShape{OMX}{MnSymbolE}{m}{n}{
    <-6>  MnSymbolE5
   <6-7>  MnSymbolE6
   <7-8>  MnSymbolE7
   <8-9>  MnSymbolE8
   <9-10> MnSymbolE9
  <10-12> MnSymbolE10
  <12->   MnSymbolE12
}{}
\DeclareFontShape{OMX}{MnSymbolE}{b}{n}{
    <-6>  MnSymbolE-Bold5
   <6-7>  MnSymbolE-Bold6
   <7-8>  MnSymbolE-Bold7
   <8-9>  MnSymbolE-Bold8
   <9-10> MnSymbolE-Bold9
  <10-12> MnSymbolE-Bold10
  <12->   MnSymbolE-Bold12
}{}

\let\llangle\@undefined
\let\rrangle\@undefined
\DeclareMathDelimiter{\llangle}{\mathopen}%
                     {MnLargeSymbols}{'164}{MnLargeSymbols}{'164}
\DeclareMathDelimiter{\rrangle}{\mathclose}%
                     {MnLargeSymbols}{'171}{MnLargeSymbols}{'171}
\makeatother

\begin{document}

\thispagestyle{empty}

\vspace*{20mm}  
\begin{center}
	{\Huge   
   Topological Constraints on Defect Dynamics}
	\\[13mm]
   {\large Andrea Antinucci$^{\symknight}$, Christian Copetti$^{\symknight}$ \\[0.6em]
        Giovanni Galati$^{\symbishop}$ and Giovanni Rizi$^{\symrook}$
        }

	\bigskip
	{\it
		 $\symknight$ Mathematical Institute, University
of Oxford, Woodstock Road, Oxford, OX2 6GG, United Kingdom \\[.6em]
    $\symbishop$  Physique Theorique et Mathematique and International Solvay Institutes
Universite Libre de Bruxelles, C.P. 231, 1050 Brussels, Belgium \\[.6em]
${\symrook}$ Institut des Hautes Etudes Scientifiques, 91440 Bures-sur-Yvette, France
	} 
\end{center}

\bigskip

 \begin{abstract}
 \noindent
Extended objects (defects) in Quantum Field Theory exhibit rich, nontrivial dynamics describing a variety of physical phenomena. These systems often involve strong coupling at long distances, where the bulk and defects interact, making analytical studies challenging. By carefully analyzing the behavior of bulk symmetries in the presence of defects, we uncover robust topological constraints on defect RG flows. Specifically, we introduce the notions of \emph{defect anomalies} and \emph{symmetry reflecting defects}, both of which are RG-invariant. Several known notions, such as higher-form symmetries, fractionalization, and projective lines, are revealed to be manifestations of defect anomalies, which also encompass novel phenomena and forbid trivial defect dynamics in the IR. Meanwhile, symmetry reflecting defects are shown to remain coupled at low energies, imposing powerful dynamical constraints. We verify our findings through concrete examples: exactly solvable defect RG flows in (1+1)d Conformal Field Theories with symmetry reflecting lines and a surface defect in (2+1)d scalar QED.
 \end{abstract}

\pagenumbering{arabic}
\setcounter{page}{1}
\setcounter{footnote}{0}
\renewcommand{\thefootnote}{\arabic{footnote}}

{\renewcommand{\baselinestretch}{.88} \parskip=0pt
\setcounter{tocdepth}{2}

\newpage

\tableofcontents}

\newpage

\section{Introduction}
Quantum field theories, as well as many-body systems, can be studied in the presence of extremely heavy external objects. In the limit where they cannot move, they are regarded as \emph{defects}, and they play a prominent role in our understanding of strongly coupled phenomena. Their importance is matched by their ubiquity: defects arise naturally, for example, in condensed matter studies as impurities \cite{Kondo:1964nea, hewson1997kondo}, and in high energy physics as disorder-type configurations \cite{Abrikosov:1956sx,Nielsen:1973cs,Tong:2005un,Shifman:2009zz} or as strings in confining theories \cite{Luscher:2002qv,Teper:2009uf,Dubovsky:2012sh,Brandt:2016xsp}, as well as in cosmology, where they may appear as extended configurations in early-universe phase transitions \cite{Durrer:2001cg}. 
Despite their widespread relevance, relatively few results are known about their dynamics compared to our understanding of the properties of local excitations.

Consequently, there has been growing effort devoted to develop novel theoretical approaches to tame defect dynamics and investigate their RG flows, extending beyond the more familiar case of line defects in (1+1) dimensions, and supersymmetric setups\footnote{For instance, computations related to BPS Wilson loop RG flows have been performed in various contexts; see, for example, \cite{Polchinski:2011im, Beccaria:2022bcr, Castiglioni:2022yes, Castiglioni:2023uus, Castiglioni:2023tci} for a representative set of references.}, see \cite{Andrei:2018die} for a review.
We will be mostly interested in defect RG flows. This means that the bulk remains a fixed CFT while the defect itself undergoes a non-trivial RG flow: bulk correlators see the lack of conformality near the defect, but asymptote to their CFT values at large distances. The breaking of conformal symmetry can arise either because the defect is described by a bare UV Lagrangian, or due to a relevant defect operator being explicitly turned on in the UV. We will refer to this setup as Defect QFT (DQFT) and, in the conformal case, as DCFT.
Such RG flows have a long history, dating back to the (1+1)d Kondo problem (see \cite{Affleck:1995ge} for a review).  

A central question in these studies is that of screening, that is whether the defect RG flow results in a nontrivial interacting bulk/defect system.  
Intuitively, screening describes a process in which the bulk degrees of freedom neutralize the interaction between the bulk and the defect at the IR critical point.   
For our purposes, a screened defect is a DQFT which at low energies factorizes into a defect and a bulk theory. There are roughly two possible fates for DQFTs in the IR:
\begin{enumerate}
    \item An \emph{unscreened} DQFT, meaning that it is described either by a gapless DCFT with a nontrivial defect OPE, or by a gapped --topological-- theory which acts nontrivially on bulk excitations.
    \item A \emph{screened} DQFT, which can either be a gapless system whose correlation functions with bulk operators factorize,\footnote{In this case, one can detect the screening by the triviality of the displacement operator.} or a gapped --topological-- object decoupled from bulk dynamics.
\end{enumerate}

In the absence of supersymmetry, existing methods for detecting screening rely either on having a free bulk theory \cite{Gorbar:2001qt,Herzog:2017xha,Lauria:2020emq,Herzog:2022jqv,Bashmakov:2024suh} or on some type of perturbative expansion, such as large $N$ \cite{Cuomo:2021rkm,Beccaria:2022bcr}, the $\epsilon$ expansion \cite{Cuomo:2021kfm, SoderbergRousu:2023zyj,Rodriguez-Gomez:2022gif,CarrenoBolla:2023sos}, or the large charge/representation of the defect \cite{Cuomo:2021cnb, Cuomo:2022xgw,Rodriguez-Gomez:2022gbz,Aharony:2022ntz,Aharony:2023amq}.\footnote{An important exception stems from bootstrap techniques applied to boundaries and defects \cite{Liendo:2012hy,Billo:2016cpy}.}  
Another significant source of constraints is encoded by defect monotonicity theorems:
the $g$-theorem \cite{Affleck:1991tk,Friedan:2003yc,Casini:2016fgb,Cuomo:2021rkm}, which applies to line defects $p=1$, and the $b$-theorem \cite{Jensen:2015swa,Wang:2020xkc,Shachar:2022fqk} for surface defects $p=2$. They state that the coefficient of the scheme-independent part of the defect free energy (the dots represent finite or scheme-dependent quantities) 
\be  
F_{\text{defect}} = - \log\left( \frac{Z_{\text{defect}}}{Z} \right) \sim \begin{cases}
 - \log(g) + \dots & p = 1 \\
 -\frac{b}{6}\log(R) + \dots & p = 2
\end{cases}  \, ,  
\ee  
is monotonically decreasing along defect RG flows.\footnote{See also \cite{Casini:2023kyj} for an entropic proof of these results.}  
 These theorems, while extremely useful, assume that the bulk theory remains conformal throughout the flow, and counterexamples are known when this assumption is violated \cite{Green:2007wr,Shachar:2024ubf}.

In this paper we study how symmetries of the bulk theory --- and their realization in the DQFTs --- can impose significant constraints on defect RG flows. Such constraints may, for example, rule out scenarios in which the IR defect is screened. Importantly, our constraints can be applied to non-conformal bulk theories, provided the bulk RG flow preserves the symmetry. The critical role of symmetry in constraining the fate of defects can already be appreciated by considering the example of line operators in gauge theories \cite{Aharony:2022ntz} charged under a one-form symmetry \cite{Gaiotto:2014kfa}, that cannot be screened in the IR, as the one-form symmetry charge is preserved along the RG flow.
This example illustrates how a nontrivial symmetry \emph{charge} in the UV can constrain the IR DQFT.\footnote{However, line of reasoning can seldom be used if $p > 1$, as higher form symmetries of degree higher than 1 are exceedingly rare in four or less spacetime dimension.} 
Our work provides new insights in this direction by identifying a class of defects, which we refer to as \emph{symmetry reflecting}, that cannot be screened during RG flows, as defined earlier.  

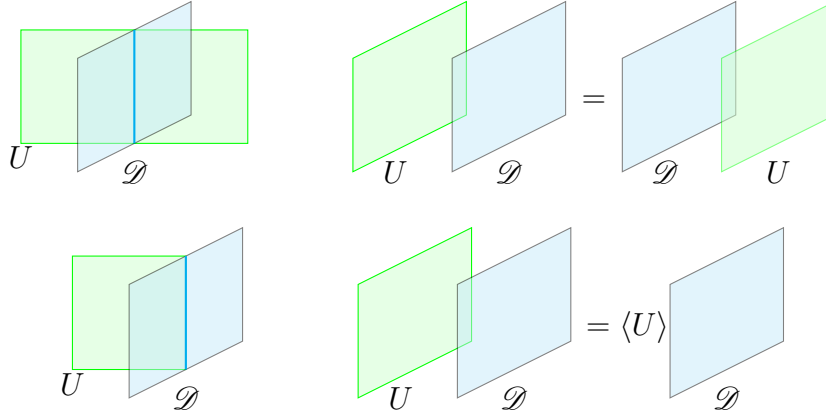
\begin{figure}[h]
    \centering
   \begin{tikzpicture}  [scale=0.75]
     \draw[color=green,fill=white!90!green] (-1,0.5) -- (3,0.5) -- (3,2.5) -- (-1,2.5) -- cycle;
     \draw[fill=white!80!cyan,opacity=0.5] (0,0) -- (2,1) -- (2,3) -- (0,2) -- cycle; 
    \node at (1,0) {$\calD$};
    \draw[thick,cyan] (1,0.5) -- (1,2.5);
    \node at (-1,0.25) {$U$};
\end{tikzpicture}  \ \ \ \ \ \ \ \ \
\begin{tikzpicture}[scale=0.75]
    \draw[color=green,fill=white!90!green] (-1.75,0) -- (0.25,1) -- (0.25,3) -- (-1.75,2) -- cycle; 
    \node at (-1,0) {$U$};
      \draw[fill=white!80!cyan,opacity=0.5] (0,0) -- (2,1) -- (2,3) -- (0,2) -- cycle; 
    \node at (1,0) {$\calD$};
    \node at (2.5,1.25) {$=$};
    \begin{scope}[shift={(4.75,0)}]
         \draw[fill=white!80!cyan,opacity=0.5] (-1.75,0) -- (0.25,1) -- (0.25,3) -- (-1.75,2) -- cycle; 
    \node at (-1,0) {$\calD$};
      \draw[color=green,fill=white!80!green,opacity=0.5] (0,0) -- (2,1) -- (2,3) -- (0,2) -- cycle; 
    \node at (1,0) {$U$};   
    \end{scope}
\end{tikzpicture} 
\\[1em]
\begin{tikzpicture}[scale=0.75]
\draw[color=green,fill=white!90!green] (-1,0.5) -- (1,0.5) -- (1,2.5) -- (-1,2.5) -- cycle;
 \draw[fill=white!80!cyan,opacity=0.5] (0,0) -- (2,1) -- (2,3) -- (0,2) -- cycle; 
    \node at (1,0) {$\calD$};
     \draw[thick,cyan] (1,0.5) -- (1,2.5);
    \node at (-1,0.25) {$U$};
\end{tikzpicture} \ \ \ \ \ \ \ \ \ \ 
\begin{tikzpicture}[scale=0.75]
     \draw[color=green,fill=white!90!green] (-1.75,0) -- (0.25,1) -- (0.25,3) -- (-1.75,2) -- cycle; 
    \node at (-1,0) {$U$};
      \draw[fill=white!80!cyan,opacity=0.5] (0,0) -- (2,1) -- (2,3) -- (0,2) -- cycle; 
    \node at (1,0) {$\calD$};
    \node at (3,1.25) {$= \langle U \rangle$};
    \begin{scope}[shift={(3.75,0)}]
     \draw[fill=white!80!cyan,opacity=0.5] (0,0) -- (2,1) -- (2,3) -- (0,2) -- cycle; 
    \node at (1,0) {$\calD$};
    \end{scope}
\end{tikzpicture}
    \caption{Above: a symmetric defect $\calD$. It can be threaded topologically by the symmetry generator $U$, implying that they commute. Below: a symmetry reflecting defect $\calD$. The topological operator $U$ can terminate on it. Equivalently $\calD$ absorbs $U$.}
    \label{fig:stronglysym}
\end{figure}

The starting point of this investigation naturally leads to the question:  
    \emph{how can a bulk symmetry be realized on extended defects?} This simple question turns out to have rather subtle answers\footnote{See \cite{Copetti:2024onh} for a SymTFT based discussion, and \cite{Konechny:2019wff,Huang:2023pyk,Choi:2023xjw,Copetti:2024rqj,Cordova:2024vsq,Cordova:2024iti,Copetti:2024dcz,Bhardwaj:2024igy,Choi:2024tri,Choi:2024wfm} for similar studies focusing on boundaries.} that we study in Sections \ref{sec: symdef} and \ref{sec: defectanomal} with a purely QFT approach.
    In Section \ref{sec: symdef} we introduce the notions of \emph{symmetric} and \emph{symmetry reflecting} defects.  It is useful to employ the characterization of symmetries in terms of topological operators \cite{Gaiotto:2014kfa}. A symmetric defect commutes with bulk topological operators implementing the symmetry, or, alternatively, bulk symmetry operator remains topological even when crossing the defect $\calD$.
A symmetry reflecting defect, instead, can absorb symmetry generators, or equivalently the symmetry generator can terminate on the defect. These two concepts are summarized in Figure \ref{fig:stronglysym}. 

In Section \ref{sec: defectanomal} we introduce the concept of a \emph{defect anomaly}. A DQFT with a $p$-dimensional defect, together with its symmetry action, is anomalous if it must be dressed with an invertible $p+1$-dimensional topological field theory ending on the defect world-volume to ensure invariance under symmetry transformations.  
Similar notions have already appeared in discussions of line operators in gauge theories under various guises \cite{Brennan:2022tyl,Delmastro:2022pfo}. 
After reviewing these examples in detail we provide a unified framework applicable to any type of defect. Defect anomalies must match along RG flows, thereby imposing strong constraints on the infrared. Note that this applies both to defect RG flows and to setups where the bulk and defect flow together. However, while defect anomalies imply a non-trivial dynamics on the defect,  they are compatible with the bulk-defect system being decoupled in the absence of background gauge fields. In fact a defect can sometimes couple to the bulk only via specific contact terms in current correlation functions, that is a reflection of a defect anomaly. 

Symmetry reflecting DQFTs, on the other hand, cannot be decoupled from the bulk in the IR, hence they are of practical importance. We provide several first-principle constructions of such defects and analyze their RG flows in Sections \ref{sec: 1+1} and \ref{sec: 3+1}. These examples are of great importance for us, as they provide solvable defect RG flows where our topological constraints can be tested.  Section \ref{sec: 1+1} is devoted to conformal line defects in $(1+1)$d CFTs. Without supersymmetry, these have been completely classified only for the $c=1/2$ Ising CFT \cite{Oshikawa:1996dj,Bachas:2013ora}.\footnote{See for example \cite{Gang:2008sz,Makabe:2017ygy} for some results in the next minimal model: $c=7/10$.} Our construction describes new defect RG flows through a generalization of the pinning field method \cite{Assaad:2013xua,ParisenToldin:2016szc,Cuomo:2021rkm}.
Section \ref{sec: 3+1}, instead, is devoted to the study of surface operators in (2+1)d $U(1)$ gauge theories with fundamental matter. These are a particularly interesting playground due to their rich net of dualities \cite{Peskin:1977kp,Dasgupta:1981zz,Seiberg:2016gmd,Karch:2016sxi} and describe critical universality classes of well-known condensed matter systems \cite{Komargodski:2017dmc,Komargodski:2017smk}. Our surface defects flow to new critical DCFTs which differ from the familiar $O(N)$ surface defect \cite{Trepanier:2023tvb,Giombi:2023dqs,Raviv-Moshe:2023yvq}.

\section{Symmetric defects}\label{sec: symdef}
In this section we will describe in detail how bulk symmetries can be realized by defects.  We will first discuss the case of continuous symmetries (generically higher-form \cite{Gaiotto:2014kfa}) and then move on to the general case.
In the study of defects, we have two well-separated concepts: intrinsic symmetries of defects --- namely, symmetries realized uniquely on the defect's world-volume, whose selection rules are not tied to those of bulk operators --- and bulk symmetries, which are unbroken by the defect and give rise to selection rules tying together bulk and defect fields. In this paper, we will primarily focus on the latter type.

\subsection{Continuous symmetries and tilt operators}
Consider a continuous 0-form symmetry $G$ --- which, for concreteness, we take to be $U(1)$ ---  with current $J$ in the presence of a $p$-dimensional defect $\calD$. The defect background will, in general, modify the bulk Ward identities to
\be \label{eq: tiltop}
d \star J = \tau \, \wedge \, \delta(\Sigma_p) \, \ \  \ \ \ \text{or} \ \ \ \ 
\partial_\mu \, J^\mu = \tau \, \delta(\Sigma_p) \, ,
\ee
where $\delta(\Sigma_p)$ is a delta function localized on the defect's world-volume, and the operator $\tau$ is called the \emph{tilt operator} \cite{Metlitski:2020cqy,Padayasi:2021sik,Drukker:2022pxk}.\footnote{See also \cite{Cuomo:2023qvp} for a recent discussion of symmetry breaking by defects and \cite{Bray:1977fvl,Ohno:1983lma} for early observations about protected boundary operators.} In a conformal setup, the dimension of $\tau$ is fixed by conformal symmetry to be
\be
\Delta_\tau = p \, ,
\ee
and thus $\tau$ is an exactly marginal operator. The presence of a nontrivial tilt operator indicates breaking of the bulk symmetry by the defect and gives rise to a defect conformal manifold.
To better understand how this happens, consider the process of sweeping the topological defect 
\be 
U_\alpha[\Gamma_{d-1}] = \exp\left(i \alpha \int _{\Gamma_{d-1}} \star J \right) \, ,
\ee
across $\calD$. Using \eqref{eq: tiltop}, we learn that the symmetry generator remains topological away from $\calD$. However, there is a contact term that arises once we cross the defect's world-volume. Taking this into account, we find that the defect action is shifted to
\be
S_\alpha = S + i \alpha \int_{\Sigma_p} \tau \, ,
\ee
describing a (finite) marginal deformation of the defect theory.\footnote{In Euliclidean signature, the factor of $i$ might seem puzzling. However, the tilt operator can also become imaginary after analytic continuation, ensuring that the overall deformation is sensible. A simple example, described in detail in \cite{Choi:2023xjw}, are Dirichlet boundary conditions for a compact scalar $X$ in $(1+1)$d. The finite deformation is $\frac{\alpha}{2 \pi} \partial_\perp X$, while in terms of the dual field it becomes a theta angle $\frac{i \alpha}{2 \pi} d \widetilde{X}$.} The conclusion, already reached in \cite{Drukker:2022pxk}, is that a nontrivial tilt operator gives rise to a defect conformal manifold. This is the correct notion of ``symmetry-breaking" by defects and has been applied in various contexts \cite{Herzog:2023dop,Copetti:2023sya,Cuomo:2023qvp}.

\subsection{Symmetric defects}
In order for a defect to respect the symmetry, the tilt operator must induce a trivial deformation, which means:
\be
\tau = d \star_p j \, . \label{eq: tilttriv} 
\ee
Here, $\star_p$ is the Hodge star operator on the defect's world-volume.\footnote{In CFT terminology, the tilt operator is a descendant field and not a primary operator.} 
The current $j$ is sometimes referred to as a \emph{defect current}. A defect hat satisfies \eqref{eq: tilttriv} is called \emph{symmetric}.
To see why, note that if equation \eqref{eq: tilttriv} is satisfied, the topological defect $U_\alpha$ can 'pass through' $\calD$ topologically, provided their intersection is dressed by $j$. More explicitly we define an improved symmetry operator:
\be
\widetilde{U}_\alpha\left[\Gamma \right] = U_\alpha\left[\Gamma\right] \, \exp\left( - i \alpha \int_{\Gamma \cap \Sigma_p} \star_p \, j \right) \, .
\ee
 Using \eqref{eq: tilttriv}, we see that we are free to deform $\Gamma$ continuously: $\widetilde{U}_\alpha$ is topological even in the presence of the defect (see Figure \ref{fig:stronglysym}).
We also see from this discussion that the action of $j$ on the defect implements the bulk $G$ symmetry. This can have a fairly rich structure on its own; in particular, we will see that defects can support 't Hooft anomalies, as discussed in Section \ref{sec: defectanomal}.\footnote{As highlighted in \cite{Cuomo:2023qvp}, the defect current $j$ vanishes in DCFT due to conformal symmetry. }

To gain familiarity consider a defect described by a QFT with $U(1)$ symmetry coupled to the bulk as:
\be
S = S_{\text{bulk}} + S_{\text{def}} + \mu \int_{\Sigma_p} \phi_+ \, \cO_+ \, .
\ee
with $\phi_+$ a bulk operator and $\cO_+$ a defect operator, both of positive charge. 
Noether's theorem implies that both the bulk and defect $U(1)$ currents, respectively $J$ and $j$, are broken, but their anti-diagonal combination is preserved:
\be \label{eq:symmetric defect}
d \star J = d \star_p j \, \wedge \delta(\Sigma_p) \, ,
\ee
which is precisely the defining equation of a symmetric defect. Similarly, a broad class of defects is defined by turning on a relevant deformation on the trivial $p$-dimensional defect using a bulk operator $\cO$. In this case, the defect is symmetric if and only if $\cO$ is a singlet deformation, and the defect current $j$ is trivial.

This definition extends to the case of higher-form symmetries \cite{Gaiotto:2014kfa}, which are implemented by a $(q+1)-$form current $J^{(q)}$. For $p > q$, equation \eqref{eq: tiltop} now becomes:
\be
d \star J^{(q)} = \tau^{(p-q)} \wedge \delta(\Sigma_p) \, ,
\ee
with $\tau_q$ being a $(p-q)-$form
. Commuting the symmetry generator
\be
U_\beta[\Gamma_{d-q-1}] = \exp\left(i \beta \int_{\Gamma_{d-q-1}} \star J^{(q)} \right) \, ,
\ee
through the defect leaves behind an integrated insertion of $\tau^{(p-q)}$. We conclude similarly that symmetric defects are those for which $\tau^{(p-q)}$ is a total derivative:
\be
\tau^{(p-q)} = d \star_p j^{(p-q-1)} \, ,
\ee
and the symmetry operator can cross a symmetric $\calD$ while remaining topological and, in particular, commutes with it:
\be
U_\beta \, \calD = \calD \, U_\beta \, .
\ee
For $p<q$, the topological operator $U_\beta$ can be passed through $\calD$ without intersecting it. Thus, $\calD$ is automatically symmetric in this case. For $p=q$, instead, the $U_\beta$ and $\calD$ defects can link each other topologically, giving rise to the well known notion of charge for higher form symmetries \cite{Gaiotto:2014kfa} (see Section \ref{sec:higher form symmetries} for a more detailed discussion on this point). 

\subsection{Symmetry reflecting defects} 
We now introduce a more stringent notion of symmetric defect, which will have immediate implications for defect RG flow: that of \emph{symmetry reflecting} defect. These are described by defects $\calD$ on which the bulk symmetry operators $U_\alpha$ are allowed to end topologically, as opposed to just passing through them (see Figure \ref{fig:stronglysym}). Such termination only makes sense if $p \geq d-q-1$, meaning the defect's world-volume dimension is larger than that of the associated symmetry defect. 
In order to assure a topological termination, we must consider modifications of the topological symmetry operator $U_\alpha$ on open manifolds $\Gamma _{d-q-1}$, with $\partial \Gamma _{d-q-1} = \gamma _{d-q-2} \subset \Sigma_p$. The bulk defect can be modified by a current $\Cop$ localized on the defect's world-volume:
\be
\widetilde{U}_\alpha [\Gamma_{d-q-1}] = \exp\left(i \alpha \int_{\Gamma_{d-q-1}} \star J \right) \, \exp\left(- i \alpha \int_{\gamma_{d-q-2}} \star_p \Cop \right) \, .
\ee
The new defect is topological provided that: 
\be \label{eq: stronglysymm}
\iota^*\left(\star J\right) =d  \star_p \, \Cop \, ,
\ee
with $\iota^*$ the pullback on the defect.\footnote{For the familiar case of codimension one defects, this is nothing but the normal component $J_\perp$ of the current. Note that, similar to the defect current $j$, the operator $\Cop$ also vanishes at the conformal fixed point.}
The discussion above makes sense only if the defect $\calD$ is already symmetric. Otherwise, integrating equation \eqref{eq: tiltop} perpendicularly to the defect shows that the current operator is discontinuous on $\calD$ and equation \eqref{eq: stronglysymm} is not sensible. Requiring a defect to be symmetry reflecting is generally a stronger condition than \eqref{eq: tiltop}, one notable exception being the case where $\Sigma_p$ is a boundary. 
In this case, denoting by $y$ the direction normal to the boundary, we can integrate equation \eqref{eq: tiltop} along $y$ to find:
\be
\iota^*\left(\star J\right) = \tau \, ,
\ee
and, if the boundary condition is symmetric, we automatically recover \eqref{eq: stronglysymm} with $\Cop=j$.

In Section \ref{ssec: realization} we develop some general constructions which give rise to UV symmetry reflecting defects. For now let us state the main point that makes this notion interesting. Consider a symmetric RG flow involving a symmetry reflecting defect $\calD$. This can also involve a bulk RG flow, with the caveat that the bulk cannot flow to a theory where the symmetry acts trivially. We ask ourselves if the endpoint of the flow $\calD_{IR}$ can be screened.
The answer is negative: consider an open defect $\widetilde{U}_\alpha$ that terminates topologically on $\calD$ in the UV. As the RG flow preserves the symmetry, the termination must remain topological at any intermediate energy scale. If $\calD_{IR}$ is screened, such topological termination must be implemented by a bulk topological termination for the defect $U_\alpha$. This is only possible if the symmetry acts trivially in the IR, which we have excluded by assumption.
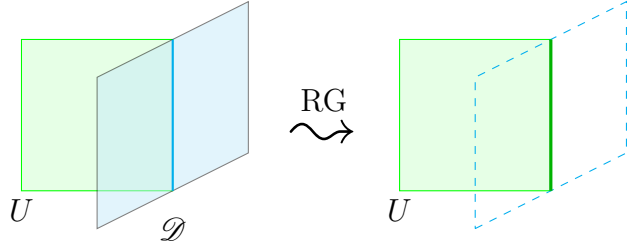
\begin{figure}
    \centering
   \begin{tikzpicture}
       \draw[color=green,fill=white!90!green] (-1,0.5) -- (1,0.5) -- (1,2.5) -- (-1,2.5) -- cycle;
 \draw[fill=white!80!cyan,opacity=0.5] (0,0) -- (2,1) -- (2,3) -- (0,2) -- cycle; 
    \node at (1,0) {$\calD$};
     \draw[thick,cyan] (1,0.5) -- (1,2.5);
    \node at (-1,0.25) {$U$};
    \node at (3,1.5) {\Huge{$\overset{\text{\normalsize{RG}}}{\leadsto}$} };
    \begin{scope}[shift={(5,0)}]
          \draw[color=green,fill=white!90!green] (-1,0.5) -- (1,0.5) -- (1,2.5) -- (-1,2.5) -- cycle;
           \draw[color=cyan,dashed] (0,0) -- (2,1) -- (2,3) -- (0,2) -- cycle; 
             \draw[very thick,black!30!green] (1,0.5) -- (1,2.5);
               
                  \node at (-1,0.25) {$U$};
    \end{scope}
   \end{tikzpicture}
    \caption{The forbidden flow screening a symmetry reflecting defect: the endpoint of the flow implies that the bulk topological operator can be terminated topologically, or, in other words, the bulk symmetry does not act faithfully.}
    \label{fig:enter-label}
\end{figure}
We thus conclude that $\calD_{IR}$ cannot decouple from the bulk. What are the possible endpoints of the flow? There are two cases:
\begin{enumerate}[a)]
    \item The defect $\calD$ flows to an IR fixed point $\calD_{IR}$ which is gapless and symmetry reflecting. This describes a nontrivial conformal defect of the theory (DCFT).
    \item On the other hand, $\calD_{IR}$ can be gapped and topological. Being symmetry reflecting implies that it must generate a nontrivial non-invertible symmetry, as invertible (group-like) symmetry defects cannot absorb other symmetry operators. Notable examples are duality defects \cite{Frohlich:2006ch,Chang:2018iay,Choi:2021kmx,Kaidi:2021xfk} and condensation defects \cite{Roumpedakis:2022aik}, which we'll encounter again in the next sections.
\end{enumerate}
Importantly, a symmetric --- as opposed to symmetry reflecting --- defect can still be screened along RG flows, as the identity defect is symmetric.\footnote{Alternatively, it is possible that the defect current $j$ flows to a trivial defect operator.} Furthermore, our definition of symmetry reflecting defects can be neatly rephrased in a clear mathematical language using the notion of (higher) module categories \cite{etingof2016tensor}: a symmetry reflecting defect is described by a module category with a single simple object.

\paragraph{Symmetric defects for general symmetries}
Our discussion can be rephrased in a straightforward manner to also include discrete symmetries, which we denote by $\cS$. These can be either invertible or non-invertible. In the absence of the notion of tilt operator, we adopt the following definitions:
\begin{enumerate}
    \item A defect $\calD$ is $\cS$-symmetric if it commutes with topological defects $U \in \cS$:
    \be
    U \, \calD = \calD \, U \, .
    \ee
    Alternatively, $U$ intersects $\calD$ in a topological manner.
    \item A defect $\calD$ is $\cS$-symmetry reflecting if it can absorb any topological defect $U \in \cS$:
    \be
    U \, \calD = \calD \, U =\langle U \rangle \, \calD \, ,
    \ee
    where $\langle U \rangle$ is the vev of the symmetry defect $U$.\footnote{For topological lines, this corresponds to the familiar concept of quantum dimension $d_U$ of $U$.} Alternatively, topological defects can terminate on $\calD$ topologically.
\end{enumerate}

Intuitively a co-dimension one symmetry reflecting defect behaves similar to a boundary for the bulk symmetry. This viewpoint also offers an alternative argument for the non-decoupling of symmetry reflecting defect: a boundary can never decouple. Moreover, we conclude that a defect $\calD$ cannot be symmetry reflecting for an anomalous symmetry of the bulk, for the same reason that anomalies forbid symmetric boundary conditions \cite{Jensen:2017eof, Thorngren:2020yht}. More generally, if a defect is symmetric under an anomalous bulk symmetry $G$, the subgroup $H\subset G$ under which is symmetry reflecting must be non-anomalous.
\subsection{Realizing symmetry reflecting defects in the UV}\label{ssec: realization}
We conclude by outlining two ways in which symmetry reflecting UV defects can be explicitly realized. We will apply these ideas in Sections \ref{sec: 1+1} and \ref{sec: 3+1}.
 \paragraph{Coupling to an anomalous defect symmetry.} If a 2d theory has a continuous symmetry with an (possibly mixed) 't Hooft anomaly we have
    \be \label{eq:perturbative anomalu}
    d \star_2 \Cop = \frac{\kappa}{2 \pi} F \, ,
    \ee
    where $\Cop$ is the current, while $F=dA$ is the field strength of background field. If the bulk has a dynamical Abelian gauge field $a$ we can couple the 2d theory by identifying $A=a$. The bulk has a $U(1)_T$ $(d-3)-$form symmetry with current $\star J=\dfrac{da}{2\pi}$, and \eqref{eq:perturbative anomalu} becomes the defining equation \eqref{eq: stronglysymm} for a defect symmetry reflecting under the bulk $U(1)$ symmetry.    This example will be discussed in great detail in Sections \ref{sec:QED3} and \ref{sec: 3+1}.
    \paragraph{Deformation of symmetry reflecting topological defects.} A second method, which can be applied effectively to line operators in $(1+1)$ dimensions, involves starting with a UV symmetry reflecting topological line $\calD_{UV}$. We consider $G=\bZ_2$ generated by $\eta$ for simplicity:
    \be
    \eta \, \times \, \calD_{UV} = \calD_{UV} \, \times \, \eta = \calD_{UV} \, .
    \ee
    The simplest examples are found in the Ising CFT: the Kramers Wannier symmetry $\cN$ and the direct sum of the group generators $1 + \eta$.
    On this defect, we turn on a relevant deformation $\phi_{\calD}$ which can be though as a deformation of the identity line \cite{Assaad:2013xua,ParisenToldin:2016szc,Cuomo:2021rkm}. 
    This deformation can be non-local, involving twisted sectors of the $G$ symmetry and needs not be bosonic either. It must, however, be both $\eta$-symmetric and commute with $\calD_{UV}$. The former constraint ensures that the deformed defect remains symmetry reflecting, while the latter that the deformation is well-defined.
    Interestingly, we show in Section \ref{sec: 1+1} that the IR fixed points of such RG flows can often be bootstrapped by a mixture of generalized symmetry action and discrete gauging operations from those of the better known pinning defects.

\section{Defect Anomalies}\label{sec: defectanomal}

It can happen that a symmetric (or symmetry reflecting) DQFT cannot be coupled to background fields in a gauge invariant fashion. This is the DQFT version of 't Hooft anomalies, and we dub them \emph{defect anomalies}. Analogously to the standard QFT case \cite{Callan:1984sa, Wess:1971yu}, the non-gauge invariance can be canceled by inflow with a higher-dimensional classical theory. This proves RG invariance of defect anomalies, making them a powerful tool. In this section we explain these concepts in detail and provide some concrete examples.
\subsection{Background fields for symmetric defect}\label{ssec: background}
If a defect is symmetric, the DQFT admits topological operators possibly crossing the defect. This can be used as the starting point to define what background fields are in DQFT, following the logic of \cite{Gaiotto:2014kfa} that a flat background gauge field describes a network of  topological operators. 
This can be made very explicit in the case of a defect symmetric under a continuous $q-$form $U(1)$ symmetry with bulk and defect currents $J$ and $j$ satisfying \eqref{eq: tilttriv}. We introduce two $(q+1)-$forms $U(1)$ gauge fields $A$ and $\cA$, living respectively in space-time and on the defect, through the minimal coupling
\begin{equation}\label{eq:couling symmetry defects}
    S[A,\cA]=S+i\int _{X_d} A\wedge \star J-i\int _{\Sigma_p} \cA \wedge \star j \ .
\end{equation}
As the two currents are not independently conserved but \eqref{eq:symmetric defect} holds, a gauge transformation $A\mapsto A+d\Lambda$, $\cA \mapsto \cA +d\lambda$ leaves the action invariant only if we identify $\iota ^*\left(\Lambda \right)=\lambda $. Hence there is no loss of generality in identifying $\cA$ with the pull-back of the $A$ on the defect:
\begin{equation}
    \cA=\iota ^*(A) \ .
\end{equation}
 A flat connection $A$ encodes the Poincare' dual of a cycle, namely $A=\alpha\, \delta(\Gamma)$. This introduces in the path integral a defect $U_\alpha = \exp\left( i \alpha \int_\Gamma \star J\right)$.
 In the presence of a symmetric defect, using \eqref{eq:couling symmetry defects} we instead insert the topological operator
 \begin{equation}
     \exp{\left(i\int _{X_d}A\wedge\left(\star J-\delta(\Sigma_p)\wedge \star j\right)\right)}=U_\alpha[\Gamma]\, \exp{\left(-i\alpha \int _{\Gamma \cap \Sigma_p}\star _p j\right)} \ .
 \end{equation}
 This makes contact with the general definition valid also for discrete symmetries, but of course \eqref{eq:couling symmetry defects} can be extended to non-flat backgrounds.

If the defect is symmetry reflecting we have additional topological operators, namely those terminating on $\calD$, and thus we can introduce more general background fields. To gain some intuition for why this is interesting, let us look at the case of a co-dimension one defect ($p=d-1$), symmetry reflecting under a $U(1)$ 0-form symmetry. The defect world-volume $\Sigma_{d-1}$ divides (at last locally) the space-time into two parts, left $X_d^{(L)}$ and right $X_d^{(R)}$, on which we  introduce two independent $U(1)$ gauge fields $A_L$ and $A_R$, together with a defect gauge field $\cA$ through the minimal coupling
\begin{equation}
    S[A_L,A_R,\cA]=S+i \int _{X_d^{(L)}} A_L\wedge \star J + \int _{X_d^{(R)}} A_R\wedge \star J -i\int _{\Sigma _{d-1}} \cA \wedge  \star \Cop  \ .
\end{equation}
Using \eqref{eq: stronglysymm} we see that a gauge transformation $A_L\mapsto A_L+d\Lambda_L, A_R\mapsto A_R+d\Lambda_R, \cA\mapsto \cA+d\lambda $ requires identifying $\iota ^*(\Lambda_L)-\iota ^*(\Lambda_R)=\lambda$, hence there is no loss of generality in assuming 
\begin{equation}
    \cA=\iota^*(A_L)-\iota ^*(A_R) \ .
\end{equation}
The important thing to notice here is that since we have two independent backgrounds $A_L, A_R$, from the point of view of the defect, the symmetry is effectively doubled 
\begin{equation}
G_{\calD}=U(1)\times U(1) \ .    
\end{equation}
This is a general fact for symmetry reflecting co-dimension one defects, regardless of whether the symmetry is continuous: we have two types of topological defects terminating on the defect, from the left of from the right, and they can be regarded as two independent symmetries. This simple observation is important to classify defect anomalies, as we will discuss shortly in Section \ref{sec: anomalies strongly symm lines} (we will then use the additional anomalies to constraint RG flows in Section \ref{sec: 1+1}). 
\subsection{Anomalies and inflow}
For a QFT with global symmetry $G$ an 't Hooft anomaly arises if the partition function $Z_{\text{QFT}}[A]$ changes by a phase under background field gauge transformation $A\mapsto A^\Lambda$,
\begin{equation}
    Z_{\text{QFT}}[A]\mapsto   Z_{\text{QFT}}[A^\Lambda]=\exp{\left(i\int _{X_d}\alpha_d(A,\Lambda) \right)}  Z_{\text{QFT}}[A]  \, ,
\end{equation}
provided that the phase cannot be removed by local counterterms. However, the anomaly $\alpha$ can be removed by coupling the QFT to an \emph{invertible} topological theory living in $(d+1)$ dimensions, with action
\be
S_{\text{inflow}} = \int_{X_{d+1}} \omega_{d+1}(A)  \, .
\ee
This is gauge invariant on closed manifolds, while we have the descent equation \cite{Wess:1971yu}
\be
\omega_{d+1}(A^\Lambda) - \omega_{d+1}(A) = d \alpha_d(A, \Lambda) \, .
\ee
This implies that choosing $\partial X_{d+1} = X_d$ we cancel the $d$-dimensional anomaly. This procedure goes under the name of \emph{anomaly inflow} \cite{Callan:1984sa}.

Consider now a DQFT symmetric or symmetry reflecting under $G$. This means, in particular, that bulk background gauge fields are coupled as discussed in \ref{ssec: background}.
In this case the failure of gauge invariance can involve an additional contribution localized on the defect:
\begin{equation}
   Z_{\text{DQFT}}[A^\Lambda,A_{\calD}^\lambda]=\exp{\left(i\int _{X_d}\alpha_{d}(A,\Lambda)+i\int _{\Sigma _p} \alpha _{p} (A,A_{\calD},\Lambda, \lambda)\right) }  Z_{\text{DQFT}}[A,A_{\calD}] \ .
\end{equation}
Here $A_{\calD}$ is the background field for any symmetry living only on the defect.  
We dub the intrinsic defect contribution $\alpha _{p} (A,A_{\calD},\Lambda, \lambda)$ the \emph{defect anomaly}. Again this is defined up to gauge variations of local counterterms on the defect.
The defect anomaly can similarly be canceled by inflow via a an invertible (p+1) dimensional TQFT, on a manifold $\Sigma_{p+1}$ with the defect's world-volume as boundary. The action now involves a topological term $\omega_{p+1}(A, A_{\calD})$ such that
\be
\omega_{p+1}(A^\Lambda, A_{\calD}^\lambda) - \omega_{p+1}(A, A_{\calD}) = d \alpha_{p}(A, A_{\calD},\Lambda, \lambda) \, .
\ee
This is the same procedure as before, but now applied to the defect.

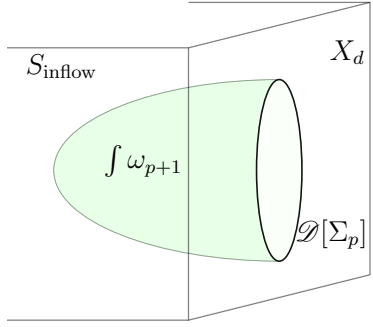
\begin{figure}[t]
$$
\scalebox{0.6}{
\begin{tikzpicture}
\draw[black, opacity=0.5] ($(0, 0) + (30:5cm and 2cm)$(P) arc
  (90:270:5cm and 2cm);
 \filldraw[white!70!green, opacity=0.3] ($(0, 0) + (30:5cm and 2cm)$(P) arc
  (90:270:5cm and 2cm);
   \filldraw[ color= white] (4.3,-1) ellipse [x radius = 0.5,y radius = 2] ;
    (90:270:0.4cm and 1.88cm);   
  \draw[line width = 1, color= black] (4.3,-1) ellipse [x radius = 0.5,y radius = 2] ;
    (90:270:0.4cm and 1.88cm)  ;  
      \filldraw[ color= white!80!green,opacity=0.1] (4.3,-1) ellipse [x radius = 0.5,y radius = 2] ;
    (90:270:0.4cm and 1.88cm) ;   
    \node[] at (5.45,-2.4) {\scalebox{1.5}{$\calD[\Sigma_p]$}};
       \node[] at (5.8,1.6) {\scalebox{1.5}{$X_d$}};
      \node[] at (1.3,-0.8) {\scalebox{1.5}{$\int \omega_{p+1}$}};
      \node[] at (-0.5,1.3) {\scalebox{1.5}{$S_{\text{inflow}}$}};
     
   \begin{scope}[shift={(-5.7,-4.3)}]
  \draw[thick,opacity=0.5] (8,0) -- (12,1) -- (12,7) -- (8,6) -- cycle;
  \draw[thick,opacity=0.5] (8,0)--(4,0);
      \draw[thick,opacity=0.5] (12,7)--(8,7);
        \draw[thick,opacity=0.5] (8,6)--(4,6);
   \end{scope}
		
  \end{tikzpicture}
 }
\hspace{1cm}
$$
\caption{A defect $\calD[\Sigma_p]$ with a defect anomaly is attached to a $(p+1)$d SPT which can be embedded into the $(d+1)$ dimensional anomaly inflow of the bulk QFT.}
\label{Fig: inflow}
\end{figure}

We will provide explicit examples of this shortly, but for the moment let us point out that there are two complementary viewpoints on the defect inflow. First, we can view the defect inflow manifold $\Sigma_{p+1}$ as embedded into the bulk $X_{d+1}$ (see Figure \ref{Fig: inflow}), so that the total anomaly inflow for the DQFT is
\begin{equation}
    S_{\text{inflow}}=\int _{X_{d+1}}\left(\omega _{d+1}(A)+\delta(\Sigma_{p+1}) \wedge \omega _{p+1}(A,A_{\calD})\right) \ .
\end{equation}

From this viewpoint, together with the discreteness of the inflow action, it follows that defect anomalies are RG invariant observables of DQFTs.

Alternatively, we may view the inflow manifold $\Sigma_{p+1}$ as embedded in physical spacetimes:
\begin{equation}
    \calD[\Sigma_p;A,A_{\calD}]=\calD[\Sigma_p] \, \exp{\left(i \int _{\Sigma_{p+1}} \omega _{p+1}(A,A_{\calD})\right) } \ .
\end{equation}
In other words, we view the $p-$dimensional defect $\calD[\Sigma_p]$ as boundary of a $(p+1)-$dimensional invertible TQFT. Even though the inflow action is now embedded in space-time, it cannot be removed by any bulk local counterterm. Thus, it remains a physical observable. For example let us consider a codimension-one defect. A defect anomaly implies that it is attached to an integral:
\be
\calD[\Sigma_{d-1},A,A_{\calD}]=\calD[\Sigma_{d-1}]\exp\left( i \int_+ \omega_{p+1}(A, A_{\calD}) \right) \, ,
\ee
on (say the right) half spacetime. A bulk counterterm of the form $- \int_{X_{d}} \omega_{d}(A, A_{\calD}) $ can be turned on at will, but it has the sole effect of redefining the defect to
\be
\calD'[\Sigma_{d-1},A,A_{\calD}]=\calD[\Sigma_{d-1}]\exp\left( - i \int_- \omega_{d}(A, A_{\calD}) \right) \, ,
\ee
preserving the inflow mechanism. This viewpoint on defect inflow helps us clarifying the fate of anomalous DQFTs upon discrete gauging operations. Suppose for example that $G$ is a finite symmetry and $\omega_{p+1}$ depends solely on the bulk gauge field $A$. Naively we would assume that gauging $G$ in the bulk would remove the defect from the theory, as it is not gauge invariant. However it actually becomes a \emph{non-genuine} $p-$dimensional defect, namely a $p-$dimensional defect living at the end of a $(p+1)-$dimensional topological defect:
\begin{equation}
\cV[\Sigma_{p+1}]=  \exp{\left(i\int _{\Sigma_{p+1}} \omega _{p+1}(a)\right)} \ ,
\end{equation}
where now $a$ is a (discrete) dynamical gauge field. Such defects have been studied extensively in several contexts \cite{Antinucci:2022vyk,Bhardwaj:2022kot,Barkeshli:2023bta}.
In technical terms, after gauging $G$ the defect $\calD$ belongs to the twisted sector of $\cV[\Sigma_{p+1}]$. This mechanism possibly gives a recipe to discover new anomalous DQFTs. 

We conclude by remarking that the full set of dynamical constraints of a given defect anomaly strongly depends on its form\footnote{This is true also for bulk anomalies. See e.g. \cite{Cordova:2019bsd, Cordova:2019jqi, Brennan:2023kpo, Apte:2022xtu, Cordova:2023bja,  Antinucci:2023ezl} for some example of how different bulk anomalies implies different constraints}. For instance while a pure anomaly for a symmetry intrinsic to the defect can obviously be matched by a completely decoupled DQFT, an anomaly also involving a faithfully acting bulk symmetry under which the defect is symmetric requires the defect to be non-decoupled, at least when the background is turned on. In this paper we discuss few specific examples, and we leave a complete account of this problem for future studies.

We will now study several examples of defect anomalies. As we will see momentarily, for line defects there are many known (but seemingly distinct) phenomena that can be understood and unified under the umbrella of defect anomalies. Examples are higher-form symmetries \cite{Gaiotto:2014kfa}, symmetry fractionalization \cite{Delmastro:2022pfo, Brennan:2022tyl}, and endable lines with projective end-points \cite{Aharony:2023amq, Brennan:2023tae}. For $p>1$ defects, on the other hand, we will discover new phenomena.

\subsection{Projective endable lines} \label{sec:projective endpoints}
One of the simplest example of defect anomalies arise in multi-flavor Abelian gauge theories. These examples have been already considered in \cite{Aharony:2023amq, Brennan:2023tae}, and here we simply reinterpreted them as defect anomalies. 
Consider a $U(1)$ gauge theory with $N_f$ complex scalars $\Phi_{i=1,...,N_f}$, and we want to study Wilson line defects
\begin{equation}\label{eq:Wilson line}
    W_q(\gamma)=\exp{\left(iq \int _\gamma a \right)} \ , \ \ \ \ q\in \bZ \ .
\end{equation}
The theory has a flavor symmetry $PSU(N_f)$, since the center $\bZ_{N_f}\subset SU(N_f)$ is part of the $U(1)$ gauge group. Equivalently, while $\Phi_i$ transform under the fundamental of $SU(N_f)$, it is not a gauge invariant operator, and the local gauge invariant operators  $\cO_{ij}=\Phi_i^\dag \Phi_j$, are in the adjoint. The Wilson line defects are symmetric under the flavor symmetry, but they have a defect anomaly if $q \neq 0 \, \text{mod}(N_f)$.

To see this we first recall that \eqref{eq:Wilson line} is only schematic because $a$ is a connection, and we can define more precisely 
\begin{equation}\label{eq:correct def W}
W_q(\gamma):=\exp{\left(iq\int _{D_2}f\right)} \ ,
\end{equation}
by viewing $\gamma$ as the boundary of a disk $D_2$, and $f=da$ is the field strength. Gauge invariance is then equivalent to the independence on $D_2$. This last property is however spoiled when we couple the bulk to a background field $A$ for the flavor symmetry $PSU(N_f)$ with non-vanishing obstruction to be lifted to $SU(N_f)$, measured by a degree-two characteristic class  $w_2(A)\in H^2(X_d,\bZ_{N_f})$. The fact that $\bZ_{N_f}\subset SU(N_f)$ is part of the gauge group means that the latter extends the flavor symmetry into $U(N_f)$, hence the dynamical gauge field is not quite a $U(1)$ gauge field, but has modified Dirac quantization condition
\begin{equation}
    \int _{\Sigma _2} \frac{f}{2\pi}=\frac{1}{N_f}\int_{\Sigma_2} w_2(A) \,  \ \text{mod}(1) \ .
\end{equation}

The non integrality of the periods of $f$ makes \eqref{eq:correct def W} dependent on $D_2$, and to correct it we must modify the definition into 
\begin{equation}
W_q(\gamma;A)=\exp{\left(iq\int _{D_2}\left(f-\frac{2\pi}{N_f}w_2(A)\right)\right)}=W_q(\gamma) \exp{\left(-\frac{2\pi i q}{N_f}\int _{D_2}w_2(A)\right)} \ .
\end{equation}
We conclude that the Wilson line of charge $q$ has a defect anomaly for the bulk flavor symmetry, with defect inflow given by
\begin{equation}
    S_\text{inflow}=-\frac{2\pi q}{N_f}\int _{\Sigma _2}w_2(A) \ .
\end{equation}
This can be viewed as a manifestation of the fact that the Wilson lines can end on non gauge invariant local operators 
that transform projectively under $PSU(N_f)$. The defect inflow action is the pull-back, through the gauge field $A$, of the projective class $\omega(q)\in H^2(B\, PSU(N_f), U(1))$.

This example has straightforward generalizations to non-Abelian gauge theories. An $SU(N_c)$ gauge theory with $N_f$ quarks in the fundamental has a flavor symmetry 
\begin{equation}
    G_F=SU(N_f)/ \bZ_{k} \ , \ \ \ \ \ k=\text{gcd}(N_c,N_f) \ .
\end{equation}
The Wilson line defects
\begin{equation}
    W_R(\gamma)=\text{Tr}_R \, P \exp{\left(i\int _\gamma a \right)}
\end{equation}
are endable on some combination of quarks, but the endpoints are generically projective, their projectivity being determined by the $N_c-$ality of the representation $\nu(R)$ as $\nu(R) \, \text{mod}(k)$. As in the Abelian example, the total group is an extension $(SU(N_c)\times SU(N_f))/ \bZ_k$ so turning on a background $A$ for the flavor symmetry requires the dynamical gauge field to be a wrongly quantized $SU(N_c)/\bZ_k$ connection with $w_2(a)=w_2(A)$. Therefore the Wilson line in representation $R$ has a defect anomaly
\begin{equation}
    S_{\text{inflow}}=-\frac{2\pi \nu(R)}{k}\int _{\Sigma_2}w_2(A) \ .
\end{equation}
By the very same mechanism, the Wilson lines also have a defect anomaly under the baryon number $U(1)_B$.

\subsection{Higher-form symmetries \& symmetry fractionalization} \label{sec:higher form symmetries}
Up to some subtlety, any $p-$dimensional defect $\calD[\Sigma_p]$ charged under a $p-$form symmetry $\bA$ can be seen as an example of a defect anomaly. Intuitively the reason is that the higher-form symmetry operator preserves the defect (it does not map it to a different defect) up to a phase, that is the action by linking.

The reason why this is subtle is that strictly speaking our definition of symmetric defect does not even apply, since generically the defect $\calD[\Sigma_p]$ and the topological operator $U_a[\Gamma_{d-p-1}]$ do not intersect. Accepting this subtlety in considering the defect symmetric, the defect anomaly follows in a straightforward manner: a topological operator linking with the defect is dually described as a flat background field that is pure gauge $B=d\lambda$, with $\lambda$ of degree $q$. Shrinking the defect corresponds to a gauge transformation, and this leaves the defect invariant up to a phase
\begin{equation}
    B\mapsto B-d\lambda  \ \ : \ \ \ \calD[\Sigma _p]\mapsto e^{i\int _{\Sigma _p} q(\lambda ) } \calD[\Sigma _p]=e^{iq(a)}\calD[\Sigma _p] \ .
\end{equation}
This phase is canceled by attaching an open surface $\exp{\left(iq \int_{\Sigma_{p+1}} B\right)}$ ending on the defect, that we can interpret as defect inflow. If $G$ is a finite symmetry, this interpretation reproduces the well known fact \cite{Aharony:2013hda} that gauging $G$ makes the charged operators into  non-genuine defects living in the twisted sector of the dual symmetry. The fact that defects charged under a higher-form symmetry cannot be screened \cite{Aharony:2022ntz, Aharony:2023amq} from this viewpoint follows from them having a defect anomaly.

One immediate consequence of this perspective is that \emph{symmetry fractionalization} \cite{Delmastro:2022pfo, Brennan:2022tyl} \footnote{See \cite{Barkeshli:2014cna} for the first introduction of this concept from the point of view of topological defects.} of a 0-form symmetry $G$ can also be seen as a defect anomaly. Symmetry fractionalization can take place whenever a 0-form symmetry $G$ co-exists with a $p-$form symmetry $\bA$. Consider for instance $p=1$ (this can be generalized for any $p$). The idea is that when two co-dimension one topological operators $U_{g_1},U_{g_2}$ of $G$ meet in co-dimension two to create $U_{g_1g_2}$, the junction can be dressed with a topological operator $V_{c(g_1,g_{2})}$ of the $1-$form symmetry, determined by a class $c\in H^{2}(BG, \bA)$. An equivalent characterization is in terms of background fields: a flat background $A\in H^1(X_d,G)$ for $G$ sources a background $B=c(A)\in H^{2}(X_d,\bA)$ for the $1-$form symmetry\footnote{In this formula we view $A$ as a map from space-time to the classifying space $BG$ of $G$ bundles, so that the pull-back $c(A):=A^*(c)$ of a class $H^{p+1}(BG,\bA)$ defines an element of the cohomology group $H^{1}(X_d,\bA)$, whose integral on $\Sigma_{2}$ is an element of the group $\bA$. Finally its pairing with the charge (representation) $q\in \bA^\vee$ gives a phase.}.  For general $p$ we have $c\in H^{p+1}(BG,\bA)$ and $B=c(A)\in H^{p+1}(X_d,\bA)$. We see that, interpreting the $p-$form symmetry itself as a defect anomaly with inflow given by $q\int B $, implies that the same defect also has a defect anomaly for $G$ with inflow action
\begin{equation}
    S_{\text{inflow}}=q\int _{\Sigma _{p+1}} c(A)  \ .
\end{equation}
The interpretation of symmetry fractionalization in terms of a topological surface attached to the defect was (for the case of line defects) already emphasized in \cite{Brennan:2022tyl}.

For 1-form symmetries ($p=1$) realized as center symmetries of some gauge group, symmetry fractionalization can be often interpreted physically as the fact that the 1-form symmetry is emergent and broken in the UV by the presence of massive matter charged under the center of the gauge group, that however transform projectively under some flavor symmetry group $G$ acting in the IR. Thus we see that this example can be connected with that of lines with projective endpoints by a bulk RG flow.
\subsection{Anomalies of symmetry reflecting lines in (1+1)d} \label{sec: anomalies strongly symm lines}
We noticed that symmetry reflecting defects allow for more general background fields, where some of the topological operators terminate on the defect. Correspondingly, there can be more defect anomalies than naively expected. A set of concrete examples, that will be used in Section \ref{sec: 1+1} to derive crucial constraints on RG flows, arises for defect lines in (1+1)d that are symmetry reflecting under a discrete finite symmetry. For a $G$-symmetric defect $\calD$ the most general defect anomaly is associated with a possible phase in the gauge transformation:
\begin{equation}
    \begin{tikzpicture}[scale=0.75]
\filldraw[color=white] (0,0) -- (2,0) -- (2, 1.5) -- (0, 1.5) -- cycle; 
    \draw[color= blue, line width = 1.5] (0,-1) node[below,black] {$\calD$} -- (0, 2);
    \draw[dashed, color=black!50!red, line width = 1] (-2, 1) -- (2, 1);
     \draw[dashed, color=black!70!green, line width = 1] (-2, 0) -- (2, 0);
    \draw[fill=black] (0, 0) circle (0.05); \draw[fill=black] (0, 1) circle (0.05);  \node[below] at (1.5, 1) {\color{black!50!red}$g$}; \node[below] at (1.5, 0) {\color{black!70!green}$h$};
      \node[right] at (2, 0.5) {$ = \ \  \chi(g,h)$};
    \begin{scope}[shift={(7, 0)}]
   \filldraw[color=white] (0,0) -- (2,0) -- (2, 1.5) -- (0, 1.5) -- cycle; 
    \draw[color= blue, line width = 1.5] (0,-1) node[below,black] {$\calD$} -- (0, 2);
    \draw[dashed, color=black!50!red, line width = 1] (-2, 0) -- (2, 0);
     \draw[dashed, color=black!70!green, line width = 1] (-2, 1) -- (2, 1);
    \draw[fill=black] (0, 0) circle (0.05); \draw[fill=black] (0, 1) circle (0.05);  \node[below] at (1.5, 0) {\color{black!50!red}$g$}; \node[below] at (1.5, 1) {\color{black!70!green}$h$};
    \end{scope}
\end{tikzpicture}
\end{equation}
This encodes a projective representation of $G$ on $\calD$ characterized by $\omega \in H^2(G, U(1))$, and here $\chi(g,h)=\frac{\omega(g,h)}{\omega(h,g)}$ determines the commutator, via an antisymmetric bicharacter on $G$. In particular there cannot be any such anomaly if $G=\bZ_n$ is a cyclic group.

Consider now the case in which $\calD$ is symmetry reflecting under $G$. 
As we have previously discussed, in the case of a symmetry reflecting interface, the symmetry acting on $\calD$ is effectively doubled $G_{\calD} = G_L \times G_R$ and this leads to a further possible anomalous phase:
\begin{equation}\label{eq:anomaly strongly symm}
    \begin{tikzpicture}[scale=0.75]
\filldraw[color=white] (0,0) -- (2,0) -- (2, 1.5) -- (0, 1.5) -- cycle; 
    \draw[color= blue, line width = 1.5] (0,-1) node[below,black] {$\calD$} -- (0, 2);
    \draw[dashed, color=black!50!red, line width = 1] (0, 1) -- (2, 1);
     \draw[dashed, color=black!70!green, line width = 1] (-2, 0) -- (0, 0);
    \draw[fill=black] (0, 0) circle (0.05); \draw[fill=black] (0, 1) circle (0.05);  \node[below] at (1.5, 1) {\color{black!50!red}$g$}; \node[below] at (-1.5, 0) {\color{black!70!green}$h$};
      \node[right] at (2, 0.5) {$ = \ \  \gamma(g,h)$};
    \begin{scope}[shift={(7, 0)}]
   \filldraw[color=white] (0,0) -- (2,0) -- (2, 1.5) -- (0, 1.5) -- cycle; 
    \draw[color= blue, line width = 1.5] (0,-1) node[below,black] {$\calD$} -- (0, 2);
    \draw[dashed, color=black!50!red, line width = 1] (0, 0) -- (2, 0);
     \draw[dashed, color=black!70!green, line width = 1] (-2, 1) -- (0, 1);
    \draw[fill=black] (0, 0) circle (0.05); \draw[fill=black] (0, 1) circle (0.05);  \node[below] at (1.5, 0) {\color{black!50!red}$g$}; \node[below] at (-1.5, 1) {\color{black!70!green}$h$};
    \end{scope}
\end{tikzpicture}
\end{equation}
As the two lines ending on $\calD$ are distinguished by which one ends from the left and which from the right, $\gamma$ is only required to be linear in both entries (hence a bicharacter) without any additional symmetry properties. This gives rise to further anomalies, which encode projective representations of the doubled group $G_{\calD}$ on the defect's world-volume.
For instance if $G=\bZ_n$ we can have nontrivial $\gamma(a,b)=\exp{\left(\frac{2\pi i r ab}{n}\right)}$, $r=0,...,n-1$.\footnote{We use $a,b,c, ...$ to denote the generators of abelian groups.} This determines a projective representation of $\bZ_n\times \bZ_n$ in which the generators $U_L$ and $U_R$ do not commute, rather 
\begin{equation}
    U_L U_R=\exp{\left(\frac{2\pi i r}{n}\right)} U_R U_L \ .
\end{equation}

A paramount example of this discussion arises if $\calD$ is a topological defect, called the \emph{duality defect} $\cN$. It fuses with invertible lines generating the Abelian group $\bA$ as
\begin{equation}\label{eq: TY fusion}
   a \times \cN = \cN \times a = \cN \ , \ \ \ \ \ \ \ \cN \times \cN =\sum _{a\in \bA}a \ .
\end{equation}
The first of these two equations tells that all the lines $a\in \bA$ can terminate on $\cN$. This is called a Tambara-Yamagami (TY) fusion category \cite{tambara1998tensor, tambara2000representations}. Among its defining data is a symmetric bicharacter $\gamma(a,b)$, which stems from commuting $a$ and $b$ lines ending on $\cN$ from the two sides.\footnote{This category is also defined by further data, see \cite{Chang:2018iay,Thorngren:2019iar} for a physics-oriented review.}
Consider for instance the (1+1)d Ising CFT that has TY symmetry for $\bA=\bZ_2$ and
\begin{equation}
    \gamma(1,1)=\exp{\left(\frac{2\pi i}{2}\right)}=-1 \ .
\end{equation}
This means that the $\bZ_2$ symmetry of the Ising model is realized projectively on the duality line. 
This discussion can be be generalized to non-topological defect by turning on symmetry preserving relevant deformations on $\cN$. The defect anomaly then puts strong constraints on the defect RG flow. We will discuss this in Section \ref{sec: 1+1}.

\subsection{A surface defect in $\text{QED}_3$}\label{sec:QED3}
We now consider a novel example of a surface defect in (2+1)d, which turns out to be both symmetry reflecting and to have a defect anomaly. Here we emphasize its defect anomaly, while in Section \ref{sec: 3+1} we analyze the dynamics. The bulk is a (2+1)d $U(1)$ gauge theory with $N_f$ massless complex scalars $\Phi_i$ of charge $1$.
It is expected that, for large enough $N_f$, the bulk theory flows in the IR to an interacting CFT. This can be rigorously proved for  $N_f\gg 1$\cite{Appelquist:1988sr}. We consider a surface defect $\calD$ obtained by coupling a (1+1)d compact boson $\phi \sim \phi +2\pi$ to the dynamical bulk $U(1)$ gauge field $a$ via the winding current $\star j_W=\frac{d\phi}{2\pi}$. After partial integration the defect action takes the form:
\begin{equation}
    S_{\text{defect}}=\int _{\Sigma_2}\left(\frac{R^2}{4\pi} d\phi \wedge \star d\phi +\frac{ik}{2\pi}\phi f\right) \ .
\end{equation}
The bulk theory has a $U(1)_T$ topological symmetry with current $\star J=\frac{ f}{2\pi}$, whose charged objects are the monopole operators. $U(1)_T$ is preserved by the defect, with vanishing defect current. Not only: the equation of motion of $\phi$ sets 
\begin{equation}
    \iota^*\left(\star J\right)=-\frac{iR^2}{2\pi k}d\star d\phi
\end{equation}
hence the defect is symmetry reflecting for $U(1)_T$ with 
\begin{equation}
    \Cop= -\frac{iR^2}{2\pi k}d\phi =\frac{1}{k}j_S \ .
\end{equation}
Here $j_S$ is the current for the shift symmetry $\phi \mapsto \phi +\theta$ that is broken by the coupling down to its $\bZ_k$ subgroup. Therefore the topological operator $U_\alpha =e^{i\alpha \int \frac{f}{2\pi}}$ of $U(1)_T$ can terminate topologically on $\calD$ by dressing the junction with $\exp{\left(i\frac{\alpha}{k}\int _\gamma \star j_S\right)}$. Notice that for $\alpha=2\pi l$ the bulk operator is trivial and we remain with $k$ topological lines on the defect generating the discrete shifts $\bZ_k \subset U(1)_S$.

There are two other symmetries in the problem. The winding $U(1)_W$ on the defect and the flavor $PSU(N_f)$ symmetry in the bulk, and they are entangled together. Naively we would conclude that $U(1)_W$ has been gauged by the coupling, but that is not quite true. While the vortex operator $V_w(x)$ with charge $w$ is not gauge invariant (it has gauge charge $kw$), it can be combined with $kw$ bulk scalar fields $\Phi_i$ to construct a gauge invariant operator. As a consequence, the gauge invariant vortices transform under the bulk flavor symmetry. 

But is $\calD$ symmetric under $PSU(N_f)$? The answer depends on $k$. Turning on a $PSU(N_f)$ background field $A$ with $w_2(A)\neq 0$ fractionalizes the fluxes of $f$:
\be
\frac{1}{2\pi} \int_{\Sigma_2} f = \frac{1}{N_f} \int_{\Sigma_2} w_2(A)  \ \ \text{mod}(1) \, ,
\ee
making the coupling ill defined unless $k$ is a multiple of $N_f$. As vortex operators dressed by bulk scalars have a non-trivial charge under the center of $SU(N_f)$, the flavor symmetry preserved by the defect is an extension of $PSU(N_f)$ by $U(1)_W$. We have the peculiar phenomenon that a bulk symmetry is not a subgroup of the symmetry of the DQFT, but is extended by a symmetry of the defect. The precise extension is a quotient 
\begin{equation}
    G_{FW}=\frac{U(1)_W\times SU(N_f)}{\bZ_{N_f}} 
\end{equation}
where the generator of $\bZ_{N_f}$ is the generator of the center in the right factor, but acts as $e^{\frac{2\pi i k}{N_f}}$ on the left hand side. The result is isomorphic to $U(N_f)$ if $N_f$ and $k$ are co-prime, but is different otherwise. Because of the quotient, a background for the flavor symmetry with $w_2(A)\neq 0$ can be turned on only if we also turn on a background $B_W$ for $U(1)_W$ with fractional fluxes 
\begin{equation}
  \int _{\Sigma_2} \frac{dB_W}{2\pi} =\frac{k}{N_f}\int _{\Sigma_2} w_2(A) \ \ \text{mod}(1) \ .
\end{equation}

If $k$ is multiple of $N_f$ the extension is trivial, and $PSU(N_f)$ acts linearly on $\calD$. In this case, however, there is a mixed defect anomaly between $PSU(N_f)$ and the discrete $\bZ_k$ shift symmetry. To see this we rewrite the coupling $\phi f$ properly using a 3d extension, and introduce the background field $B_S$ for $\bZ_k$:\footnote{we regard this as a flat $U(1)$ background with periods multiple of $\frac{2\pi}{k}$}
\begin{equation}
    \int _{\Sigma_3} \frac{ik}{2\pi}(d\phi-B_S)\wedge f \ .
\end{equation}
In the presence of a nonzero $w_2(A)$, this coupling is not independent on the 3d extension $\Sigma_3$ unless we modify it by adding
\be
\frac{ik}{ N_f} \int_{\Sigma_3} B_S\wedge w_2(A) \, .
\ee
This is an example of defect anomaly inflow. We can rewrite this in a more standard form setting $B_S=\frac{2\pi}{k}b_S$, with $b_S$ a discrete gauge field valued in $\bZ_k$:
\begin{equation}
    S_{\text{inflow}}=\frac{2\pi i }{N_f}\int _{\Sigma_3}b_S\cup w_2(A) \ .
\end{equation}

\section{Symmetry reflecting conformal lines in $(1+1)$d CFTs}\label{sec: 1+1}
In this section we illustrate a general construction of symmetry reflecting lines in (1+1)d CFTs, and we also explore specific examples of defect RG flows. The construction requires the basics of the formalism of topological defect lines in (1+1)d CFTs, for which we refer the reader to  \cite{Chang:2018iay, Thorngren:2019iar, Komargodski:2020mxz}\footnote{See \cite{Verlinde:1988sn, Petkova:2000ip, Fuchs:2002cm, Frohlich:2003hm, Frohlich:2004ef, Frohlich:2006ch, Frohlich:2009gb, Carqueville:2012dk} for earlier discussion that do not use the language of generalized symmetries.}.
\subsection{General construction}
Consider a (1+1)d CFT, symmetric under a 0-form symmetry $G$. Here we will only consider $G$ to be a finite group, with $L_g$ the corresponding topological lines. Continuous groups and non-invertible symmetries can be studied along the same lines.
We first expand on \ref{ssec: realization} and describe constructions of $G$-symmetry reflecting line defects. Specifically, let us consider a topological line $\calD_{\text{UV}}$ which is also $G$-symmetry reflecting
\be
L_g \times \calD_{\text{UV}} = \calD_{\text{UV}} = \calD_{\text{UV}} \times L_g \quad,\quad g\in G \, .
\ee
We do not assume $\calD_{\text{UV}}$ to be be a simple topological line of the theory\footnote{Simple lines are those that cannot be expressed as a direct sum of other line operators with positive integer coefficients.}, and instead include the possibility for $\calD_{\text{UV}}$ to be a sum of lines. By construction $\calD_{\text{UV}}$ must be a non-invertible topological line, as invertible lines cannot absorb non-trivial lines. When $\calD_{\text{UV}}$ is a simple line, it generates an additional, non-invertible, symmetry of the CFT. Alternatively one can always construct a non-simple but symmetry reflecting topological line as
\be\label{eq: non-simple symmetry reflecting line}
 \calD_{\text{UV}} = \sum\limits_{g\in G} L_g\,.
\ee

Regardless of the way we construct $\calD_{\text{UV}}$, we would like to turn on a relevant pinning deformation $\phi_{\calD}$ on it while preserving its symmetry reflecting nature. This deformation will produce a non-topological defect, defined as
\be
\calD_{\lambda}[\gamma] := \calD_{\text{UV}}[\gamma] \, \exp{\left(\lambda \int  _{\gamma} \phi_{\calD}\right)}
\ee
where $\phi_{\calD}$ is a defect operator and the integral is over the defect's world-line $\gamma$.\footnote{For ease of notation, we will suppress the coupling to the defect's metric.} Such deformation generically triggers a defect RG flow, which will end in an IR defect denoted by $\calD_{\text{IR}}$ and which we aim to constrain.\footnote{A similar construction was proposed in \cite{Chang:2018iay} and related constraints were put forward in \cite{Copetti:2024onh,Choi:2024tri} from a SymTFT perspective.}

By radial quantization and state-operator correspondence, the possible local defect deformations $\phi_{\calD}$ are identified with states in the Hilbert space twisted by $\calD_{\text{UV}}\times \calD^{\dagger}_{\text{UV}}$, where $\calD^{\dagger}$ is the orientation reversal of $\calD$:\footnote{Recall that, for a line defect $L$, its twisted Hilbert space $\cH_{L}$ is obtained by quantization on the cylinder with $L$ stretched along the time direction. This describes \emph{non-local} operators, which are endpoints of the line $L$. See \cite{Chang:2018iay} for the modern perspective and \cite{Petkova:2000ip} for the original treatment.}
\be
\begin{tikzpicture}[baseline={(0,1)}]
    \draw[blue,thick,->] (0,0) node[below,black] {$\calD_{UV}$} -- (0,0.15); \draw[blue,thick,->] (0,0.15) -- (0,1.85);  \draw[blue,thick] (0,1.85) -- (0,2);
    \draw[fill=black] (0,1)  circle (0.05);
    \node at (0.3,1) {\small$\phi_{\calD}$};
    \draw[dashed] (0,1) circle (0.6);
    \end{tikzpicture}
    \ \ \underset{\text{Radial quant.}}{=} \ \
    \begin{tikzpicture}[baseline={(0,1)}]
        \draw (0,0) ellipse (0.5 and 0.25);
         \draw (0,1.75) ellipse (0.5 and 0.25);
         \draw (-0.5,0)-- (-0.5,1.75);    \draw (0.5,0)-- (0.5,1.75);
         \draw[blue,thick,->] ({0.5*cos(45)},{0.25*sin(45)}) -- ($(0,0.875) + ({0.5*cos(45)},{0.25*sin(45)}) $) ;
        \draw[blue,thick] ({0.5*cos(45)},{0.25*sin(45)}) node[right] {$\calD_{UV}$} -- ($(0,1.75) + ({0.5*cos(45)},{0.25*sin(45)}) $) ;
        \draw[blue,thick,-<] ({0.5*cos(-135)},{0.25*sin(-135)}) node[below] {$\calD_{UV}^\dagger$} -- ($(0,0.875) + ({0.5*cos(-135)},{0.25*sin(-135)}) $) ;
        \draw[blue,thick] ({0.5*cos(-135)},{0.25*sin(-135)}) -- ($(0,1.75) + ({0.5*cos(-135)},{0.25*sin(-135)}) $) ;
\end{tikzpicture} \, .
\ee
If $\calD_{\text{UV}}$ was invertible,  $\calD_{\text{UV}}\times  \calD ^\dag _{\text{UV}}=1$ and the possible deformations would have been local bulk operators. On the other-hand, for symmetry reflecting defects, we can distinguish between two classes of deformations:
\begin{enumerate}
    \item \emph{Local deformations.} Triggered by a bulk local operator $\phi_{\calD}$ which commutes with $\calD_{\text{UV}}$.\footnote{If $\phi_{\calD}$ does not commute with $\calD_{\text{UV}}$, it has a discontinuity along its world-line, making the pinning deformation ill-defined.}
    If $\phi_{\calD}$ is uncharged with respect to the symmetry $G$, the deformed defect $\calD_{\lambda}$ is still symmetric. This is the only type of deformation available for invertible topological lines.
   As the deformation was induced by a local bulk operator, the defect can be factorized
    \be
    \calD_{\lambda}[\gamma] := \calD_{\text{UV}}[\gamma] \, \exp\left(\lambda \int_{\gamma_R} \phi_{\calD}\right),
    \ee
    where now the integral is performed on a curve $\gamma_R$ slightly to the right (for instance) of the defect's world-line $\gamma$. Consequently, the dynamics is entirely disentangled from the choice of $\calD_{\text{UV}}$ and all the information can be recovered by performing fusion of $\calD_{\text{UV}}$. We will not be interested in such flows in the following.

 \item \emph{Twisted deformations.} In the case of  $G$-symmetry reflecting defects, we can also allow for deformations triggered by non-local operators $\phi_{\calD}$. These are in one-to-one correspondence with linear superpositions of states in various twisted sectors:
 \be 
 \vert \phi_{\calD} \rangle = \sum_{L \, \in \, \calD \times \calD ^\dag } v_{\calD}^L \vert \phi^L \rangle \, , \ \ \  \ \ \vert \phi^L \rangle \, \in \, \cH_L \, . 
 \ee
 The operators $\phi^L$ must share the same conformal dimension and spin and the coefficients $v_{\calD}^g$ must ensure that:
 \be
\begin{tikzpicture}[baseline={(0,1)}]
   \draw[thick, blue] (0,0) node[below] {$\calD_{\text{UV}}$} -- (0,2);  
   \draw[dashed, very thick] (-1,1) -- (0,1);  
    \draw[fill=black] (0,1) circle (0.05);
    \draw[fill=black] (-1,1)  node[left] {$\phi_{\calD}$} circle (0.05);
\end{tikzpicture}
=
\begin{tikzpicture}[baseline={(0,1)}]
      \draw[thick, blue] (0,0) node[below] {$\calD_{\text{UV}}$} -- (0,2);  
   \draw[dashed, very thick] (1,1) -- (0,1);  
    \draw[fill=black] (0,1) circle (0.05);
    \draw[fill=black] (1,1)  node[right] {$\phi_{\calD}$} circle (0.05);
\end{tikzpicture}
 \ee
 i.e. the non-local operator must commutes with $\calD_{\text{UV}}$. If furthermore $\phi_{\calD}$ is uncharged under $G$, we conclude that the deformed operator $\calD_{\lambda}$ is $G$-symmetry reflecting and it cannot be entirely screened in the IR. 
\end{enumerate}
The explicit form of this commutation relation involves the matrix elements of the \emph{Tube algebra} \cite{Chang:2018iay,Aasen:2020jwb,Lin:2022dhv}, which encodes the symmetry action on twisted sectors.\footnote{While a lot is known about the general theory of such representations, explicit realizations of its matrix elements are seldom derived, see e.g. \cite{Bhardwaj:2023idu}. } 

Contrary to the previous case, twisted deformations of topological defects do not reduce, along the RG flow, to a tensor product product between a topological defect and a pinning defect. Therefore, the prediction coming from the symmetry reflecting nature of $\calD_{\lambda}$ is genuinely new.
 Let us emphasize that since $g\left( \calD_{\text{UV}} \right) > 1$\footnote{If $g_{\text{UV}} = 1$, the defect $\calD_{\text{UV}}$ cannot be symmetry reflecting, as it is invertible.}, it is not guaranteed that the IR fixed point is conformal (as opposed to topological). However, flows to other topological lines of the CFT can often be excluded by defect anomaly matching. We'll now discuss examples of such scenarios.
\subsection{$\bZ_2$-symmetry reflecting line defects in minimal models}
 We start by looking at twisted deformations of non-simple symmetry reflecting topological lines. For concreteness, let us consider the discrete series of minimal models $\cM(m+1,m)$, with $m \geq 3$ and diagonal partition function. These are unitary CFTs with central charge
\begin{equation}
    c = 1 - \frac{6}{m(m+1)}
\end{equation}
and a finite set of primary operators with conformal weights
\begin{equation}
    h_{r,s}= \frac{(m(r-s)+r)^2-1}{4m(m+1)}\,,
\end{equation}
where $r,s$ are restricted to the range $1\leq r< m$, $1\leq s < m+1$ with the identification $(r,s) \sim (m+1-r, m-s)$. In diagonal theories, for each primary operator $\phi_{r,s}$ we have a topological Verlinde line $L_{r,s}$, which fuses according to the same fusion rules of the corresponding primary operators. It is then easy (see e.g. \cite{DiFrancesco:1997nk}) to see that for any $m$, we have a $\bZ_2$ 0-form symmetry generated by the topological line $L_{1, m}\equiv L_{m-1, 1}$ which acts on other lines as
\begin{equation}
    L_{1,m} \times L_{r,s} = L_{r, m-s+1}\, .
\end{equation}
We can then define a $\bZ_2$-symmetry reflecting topological line as ($L_{1,1}$ is the identity line):
\be
\calD_{\text{UV}} := L_{1,1}+ L_{1,m}\,.
\ee

As previously explained, to construct a non-topological symmetry reflecting line we should look for relevant operators living in the $\bZ_2$ twisted sector which are uncharged under $\bZ_2$. This automatically implies that they commute with $\calD_{\text{UV}}$ and that their deformation preserves its symmetry reflecting nature.
This can be achieved by looking at how the $\bZ_2$ topological line $L_{1,m}$ acts on the $\bZ_2$ twisted partition function: a $\bZ_2$ twisted operator is $\bZ_2$-even if and only if it is a boson \cite{Chang:2018iay}. 
We find that there is always (at least) one operator compatible with the above conditions, thus defining a non-trivial symmetry reflecting line in all the diagonal minimal models (see Appendix \ref{app: Z2 minimal models} for the computation).
Following the general constraints of symmetry reflecting defects, such lines can flow either to a conformal line defect or to a non-trivially acting topological line.

\paragraph{Ising model.} In the case of $m=3$, the corresponding minimal model is the  Ising CFT. In this case the entire set of conformal line defects is known \cite{Oshikawa:1996dj}. Therefore we can explicitly compute the defect RG flow and check that IR fixed point of the symmetry reflecting defect is  non-trivial.  

The Ising CFT posses three topological Verlinde lines, usually denoted as $\{1,\eta,\cN\}$ with fusion rules
\be
\eta^2 =1\quad,\quad \eta\times \cN =\cN \times \eta= \cN \quad,\quad \cN \times \cN = 1+\eta\,. 
\ee
Thus, there are two natural candidates for a $\bZ_2$-symmetry reflecting UV defect: the duality defect $\cN$ and $1 + \eta$. 
Let us first consider $\cN$. A general $\phi_{\calD}$ must be a linear sum of $\bZ_2$ twisted and untwisted primary operators. The latter are $\unit$, $\epsilon,$ and $\sigma$, while the former are given by $\psi^+$, $\psi^-$ and $\mu$. The only $\bZ_2$-even operator in the twisted sector is the disorder operator $\mu$. However since by KW duality
\be
\begin{tikzpicture}[baseline={(0,1)}]
    \draw[thick,blue] (0,0) node[below] {$\cN$} -- (0,2);
    \draw[dashed] (0,1) -- (-1,1); \node[above] at (-0.5,1) {$\eta$};
    \draw[fill=black] (-1,1) node[left] {$\mu$} circle (0.05);
\end{tikzpicture} = 
\begin{tikzpicture}[baseline={(0,1)}]
 \draw[thick,blue] (0,0) node[below] {$\cN$} -- (0,2);
    \draw[fill=black] (1,1) node[right] {$\sigma$} circle (0.05);
\end{tikzpicture} \, ,
\ee
and $\sigma$ is $\bZ_2$-odd,
there is no relevant linear combination of order and disorder operators which is invariant under both $\bZ_2$ and $\cN$: no symmetry reflecting deformation of this line can be constructed.

We then consider the second candidate, $\calD_{UV}= 1 + \eta$. Following the previous discussion, and results of Appendix \ref{app: Z2 minimal models}, it is clear that the $\mu$ deformation is both consistent and $\bZ_2$-symmetric, we then study
\be
\calD_{\lambda} = (1+\eta) \, \exp{\left(\lambda \int\mu\right)}\, .
\ee
Since $\mu$ has conformal dimension $1/8$, this is a relevant, symmetry reflecting deformation of $\calD_{UV}$. As $g\left( \calD_{\text{UV}} \right) = 2 > 1$ we cannot immediately argue that this defect must be conformal in the IR. The only other possibility being a flow to the $\cN$ defect.
To exclude this consider the $\bZ_2$ defect anomaly:
\bea
\begin{tikzpicture}[baseline={(0,0.75)}]
\filldraw[color=white] (0,0) -- (2,0) -- (2, 1.5) -- (0, 1.5) -- cycle; 
    \draw[color= blue, line width = 1.5] (0,0) -- (0, 1.5);
    \draw[dashed] (-1.5, 0.5) -- (0, 0.5); \draw[dashed] (0, 1) -- (2, 1); \draw[fill=black] (0, 0.5) circle (0.05); \draw[fill=black] (0, 1) circle (0.05); \node[above] at (-1, 0.5) {$\eta$}; \node[below] at (1.5, 1) {$\eta$};
   \end{tikzpicture}
   \  = \  \gamma \ \ \
\begin{tikzpicture}[baseline={(0,0.75)}]
    \filldraw[color=white] (0,0) -- (2,0) -- (2, 1.5) -- (0, 1.5) -- cycle;
         \draw[color= blue, line width = 1.5] (0,0) -- (0, 1.5);
    \draw[dashed] (-1.5, 1) -- (0,1); \draw[dashed] (0, 0.5) -- (2, 0.5); \draw[fill=black] (0, 0.5) circle (0.05); \draw[fill=black] (0, 1) circle (0.05); \node[below] at (-1, 1) {$\eta$}; \node[above] at (1.5, 0.5) {$\eta$};
\end{tikzpicture} \, , \ \ \  \ \ \ \gamma = \pm 1 \, .
\eea
It is well known \cite{Chang:2018iay} that $\cN$ has $\gamma=-1$, while $1 + \eta$ has $\gamma = 1$: defect anomaly matching thus forbids a $\bZ_2$-symmetric RG flow between them.
This implies that $\calD_{\text{IR}}$ must be a conformal line of the theory.

To test this prediction in this simple case, let us now \emph{bootstrap} explicitly the form $\calD_{\text{IR}}$. This defect satisfies the conditions
\bea\label{eq: Bootstrap Ising}
&\eta \times \calD_{\text{IR}} = \calD_{\text{IR}} \times \eta = \calD_{\text{IR}} \\
 &\cN \times \calD_{\text{IR}} = \left(\calD_{\text{IR}}^{\sigma,+}+\calD_{\text{IR}}^{\sigma,-}\right) \times \cN \, ,
\eea
where the first condition signals the symmetry reflecting nature of $\calD_{\lambda}$ while the second one can be proven using the definition $\calD_{\lambda}$ and explicitly commuting $\cN$ through it. Here $\calD_{\text{IR}}^{\sigma,\pm}$ are the IR fixed points of
\be
\calD^{\sigma,\pm} = \exp{\left(\pm \lambda \int\sigma \right)}\,.
\ee
Which are known to flow to \cite{yifantalk}\footnote{These flows can be nicely understood by considering a thin slab deformed by the bulk operator $\sigma$ \cite{yifantoapp}.}
\be \label{eq: IRconditions}
\calD_{\text{IR}}^{\sigma,+}=\vert + \rangle \langle + \vert \ \ \ \text{and} \ \ \ \calD_{\text{IR}}^{\sigma,-}=\vert - \rangle \langle - \vert \, ,
\ee
where $\vert x \rangle$ is a Cardy state \cite{Cardy:1989ir}. The Ising CFT has three such states: $\vert f \rangle$ and $\vert \pm \rangle$, which are respectively a singlet and a doublet under the $\bZ_2$ action \cite{Cardy:1989ir,Oshikawa:1996dj}.
Defects of the form $\vert x \rangle \langle y \vert$ are obtained by collapsing together two conformal boundary conditions. Pictorially
\bea
\begin{tikzpicture}
\filldraw[color=white] (0,0) -- (2,0) -- (2, 1.5) -- (0, 1.5) -- cycle; 
    \draw[color= blue, line width = 1.5] (0,0) -- (0, 1.5) node[above]{$\vert x \rangle \langle y \vert$};
    \node[right] at (1., 0.75) {$ = \quad \lim\limits_{\epsilon\to 0}$};
    \begin{scope}[shift={(4., 0)}]
    \filldraw[color=white] (0,0) -- (2,0) -- (2, 1.5) -- (0, 1.5) -- cycle;
    \draw[color= blue, line width = 1.5] (0,0) -- (0, 1.5) node[above]{$\vert x \rangle $};
    \draw[color= blue, line width = 1.5] (1,0) -- (1, 1.5) node[above]{$\langle y \vert$};
    \filldraw[color=white!50!black, opacity=0.2] (0,0) -- (1,0) -- (1, 1.5) -- (0, 1.5) -- cycle;
    \draw[<-] (0,-0.2) -- (0.3, -0.2) ;
    \draw[->] (0.7,-0.2) -- (1, -0.2) ;
    \node[] at (0.5, -0.2) {$ \epsilon$};
    \end{scope}
\end{tikzpicture}
\eea
Using $\cN \vert \pm \rangle = \vert f \rangle$ and imposing \eqref{eq: Bootstrap Ising}, we find a unique solution to the bootstrap problem:
\be
\calD_{\text{IR}}= \DEF{f}{f} \, .
\ee
Since $g\left(\calD_{\text{IR}}\right)=1$, the $g$-theorem is satisfied. This is a conformal --but factorized-- defect.
\subsection{Symmetry reflecting line defects at $c=1$}
We now turn our attention to a slightly different example, where $\calD_{\text{UV}}$ is a \emph{simple} symmetry reflecting line operator, describing a non-invertible symmetry of the CFT. 
A natural candidate for $\calD_{\text{UV}}$ is the duality defect $\cN$ of a Tambara Yamagami fusion category TY$(\bA)_{\gamma,\epsilon}$, introduced in Section \ref{sec: defectanomal}.
From the fusion rules \eqref{eq: TY fusion}, it is clear that the topological line $\cN$ is $\bA$-symmetry reflecting. 
The corresponding tube algebra, however, prohibits the existence of a non-genuine local operator that simultaneously commutes with both $\cN$ and $\bA$: uncharged twisted operators are always mapped by $\cN$ into local operators with non-vanishing global $\abA$ charge. 
The situation is ameliorated if we focus on non-genuine operators that commute with both $\cN$ and only a subgroup $\bB \subset \bA$.

The simplest example is that of $\abA = \bZ_2 \times \bZ_2$. For concreteness we will focus on the anomaly-free cases of $\text{Rep}(D_8) = \text{TY}(\bZ_2 \times \bZ_2)_{\gamma_o, +}$ and $\text{Rep}(H_8)= \text{TY}(\bZ_2 \times \bZ_2)_{\gamma_d, +}$.\footnote{$\gamma_d = \left( \begin{array}{cc}
    1 & 0 \\
    0  &  1
\end{array} \right)$ and $ \gamma_o = \left( \begin{array}{cc}
    0 & 1 \\
    1  & 0
\end{array} \right)$ are symmetric bicharacters on $\abA \times \abA$, which define the action of the duality line $\cN$.}
These symmetries are realized, for example, on the orbifold branch of the $c=1$ compact boson \cite{Thorngren:2021yso}.\footnote{We will follow the notation of \cite{Thorngren:2021yso} and fix a normalization for the radius $R$ such that $R=\sqrt{2}$ is the $SU(2)_1$ point and $R_{\text{orb}}=R/2$.}
The Ising$^2$ CFT lies within the orbifold branch at $R_{\text{orb}}=1$. The reader is referred to \cite{Thorngren:2021yso} for further details. 

Let us outline the representation theory required for our purposes. The group $\bZ_2 \times \bZ_2$ --- whose elements we label by $(a, b)$ with $a, b = 0, 1$ --- contains three $\bZ_2$ subgroups:
\be
\bZ_2^L, \quad \bZ_2^R, \quad \text{and} \quad \bZ_2^D ,
\ee
generated, respectively, by $(1, 0)$, $(0, 1)$, and $(1, 1)$. The relevant fusion rules are:
\be
\eta \, \cN =  \cN \eta = \cN \, , \ \ \ \eta \in \left\{ L, R, D \right\} \, , \ \ \ \ \ \
\cN^2  =1 + \eta_L + \eta_R + \eta_D \, .
\ee
Local operators charged under the $\bZ_2 \times \bZ_2$ symmetry, which we denote by $\sigma_{L,R,D}$, have charges $(q_{L},q_{R},q_{D}) = \{(-1,1,-1),(1,-1,-1),(-1,-1,1)\}$ respectively. Under the action of $\cN$, $\sigma_{L,R,D}$ are mapped to disorder operators, which we denote $\mu_{L,R,D} \, \in \, \cH_{L,R,D}$, respectively. The precise map depends on the choice of bicharacter $\gamma$ and is extracted from the consistency condition
\be
\begin{tikzpicture}[baseline={(0,1)}]
    \draw[blue] (0,0) node[below] {$\cN$} -- (0,2);
    \draw[dashed] (-1,1) circle (0.5); \node[left] at (-1.5,1) {$\eta$};
    \draw[fill=black] (-1, 1) node[below] {$q$} circle (0.05);
\end{tikzpicture}
=
\begin{tikzpicture}[baseline={(0,1)}]
    \draw[blue] (0,0) node[below] {$\cN$} -- (0,2);
    \draw[dashed] (0,0.5) arc (-90:-270:0.5 and 0.5); \node[left] at (-0.5,1) {$\eta$};
    \draw[fill=black] (1, 1) node[below] {$\mu$} circle (0.05);
    \draw[dashed] (0,1) -- (1,1); \node[above] at (0.5,1) {$\eta'$};
\end{tikzpicture} \ \ \ \ \ \ \ \ q(\eta) = \gamma(\eta, \, \eta') \, , \ \ \ \eta, \, \eta' \, \in {L,R,D} \, .
\ee
The two choices of $\gamma$ lead to the following $\cN$ action\footnote{The overall coefficient is unphysical and can be set to one by normalizing the disorder operators $\mu$.} 
\bea
\gamma_d: \,  \sigma_L \longrightarrow \mu_L \, , \ \ \ \sigma_R \longrightarrow \mu_R , \, \ \ \ \sigma_D \longrightarrow \mu_D \, , \\
\gamma_o: \,  \sigma_L \longrightarrow \mu_R \, , \ \ \ \sigma_R \longrightarrow \mu_L , \, \ \ \ \sigma_D \longrightarrow \mu_D \, .
\eea
Given any $\bZ_2$ subgroup, we can construct a symmetry reflecting $\bZ_2$ defect by the following deformations of $\cN$:
\be
\begin{array}{|c|c|c|c|} \hline
             & \bZ_2^L & \bZ_2^R & \bZ_2^D \\ \hline
   \gamma_d  &  \sigma_L + \mu_L & \sigma_R + \mu_R & \sigma_D + \mu_D \\
    \gamma_o &   \sigma_L + \mu_R & \sigma_R + \mu_L & \sigma_D + \mu_D \\ \hline
\end{array}
\ee
When the deformation is relevant it triggers a non-trivial defect RG flow that must end in a non-decoupled defect.

We can apply this idea to $c=1$ on the orbifold branch, that posses the TY$(\bZ_2\times \bZ_2)$ symmetry, and the deformation discussed above are indeed relevant. The discussion of the symmetry action follows \cite{Thorngren:2021yso} closely and we refer the interested reader for a pedagogical exposition. Local primary operators on the orbifold branch are the charge conjugation invariant vertex operators:
\be
V^{+}_{n,w} = \frac{V_{n,w} + V_{-n,-w}}{\sqrt{2}} \, ,
\ee
and the $C$-twisted sectors $\sigma_L, \, \sigma_R, \, \tau_L, \, \tau_R$ of the free boson CFT.
On these operators $\bZ_2^L\times \bZ_2^R$ acts as follows:
\be
\begin{array}{|c|c|c|c|} \hline
     & \bZ_2^L & \bZ_2^R & \bZ_2^D  \\ \hline
 V_{n,w}^+    & (-1)^n & (-1)^n & 1 \\
 (\sigma_L, \, \sigma_R)    & (-1,1) & (1,-1) & (-1,-1) \\ \hline
\end{array}
\ee
the action on $(\tau_L, \tau_R)$ being the same as on the sigmas. $\bZ_2^D$ is identified with the quantum symmetry of the orbifold branch. In Ising$^2$ notation, $\bZ_2^L$ and $\bZ_2^R$ are the $\bZ_2$ symmetries of the two decoupled Ising factors, while $\cN = \cN_1 \cN_2$ or $s_{12} \cN_1 \cN_2$ --where $s_{12}$ is the outer automorphism exchanging the two models-- depending on whether we study Rep$(H_8)$ or Rep$(D_8)$.
In the free boson variables --- for sufficiently large radius --- the lightest order-disorder pair is
\be
\sigma_D = \cos(X) \, , \ \ \ \mu_D = \sin(X) \, ,
\ee
which are relevant defect deformations as long as $R>1$. We are thus interested in the flow:
\be
\calD_\lambda = \cN \, \exp \left( \lambda \int \cos \left(X - \frac{\pi}{4} \right) \right) \, .
\ee
The broken $\bZ_2^{L,R}$ symmetries shift the scalar by $\pi$ and flip the sign of the deformation.
On the circle branch, the defect deformation
$\lambda \cos\left(X - \theta\right)$ pins the scalar on the two sides to a Dirichlet boundary condition:
\be \label{eq: pinfreefield}
\big| D , \, \pi + \theta \big\rangle \big\langle D, \, \pi+\theta \big| \, , \ \ \ \lambda> 0 \, \ \ \ \ \ \ \ \ \ \ \ \ 
\big| D , \,  \theta \big\rangle \big\langle D, \, \theta \big| \, , \ \ \ \lambda< 0 \, .
\ee
That have $g\left(D, \theta \right) = 1/\sqrt{R}$.\footnote{This follows from the 
expansion in terms of Ishibashi states:
\be
|D, \theta \rangle = \frac{1}{\sqrt{R}} \sum_{n=-\infty}^{\infty} e^{i \theta n} |n,0\rrangle \, .
\ee }
We once again bootstrap the endpoint of the defect RG flow. First, we map the problem on the circle branch. It turns out that our symmetry reflecting defect here reads:
\be
\calD^{\text{circle}}_{\lambda} = C U_{\frac{\pi}{2}} \, \exp\left( \lambda \int \cos\left(X - \frac{\pi}{4}\right) \right) + U_{\frac{\pi}{2}} C \, \exp\left( \lambda \int \cos\left(X + \frac{\pi}{4}\right) \right) \, ,
\ee
where $U_{\frac{\pi}{2}}$ is a $\bZ_4$ shift symmetry generator. Using \eqref{eq: pinfreefield} one finds:
\be
\calD^{\text{circle}}_{IR} = \begin{cases}
\big|D, \frac{3\pi}{4} \big\rangle \big\langle D, \frac{3\pi}{4}\big|   + \big|D, -\frac{3\pi}{4} \big\rangle \big\langle D, - \frac{3\pi}{4}\big|   , \, &\lambda > 0 \vspace{0.3cm}\\
\big|D, \frac{\pi}{4} \big\rangle \big\langle D, \frac{\pi}{4}\big|   + \big|D, -\frac{\pi}{4} \big\rangle \big\langle D, - \frac{\pi}{4}\big|   , \, &\lambda < 0
\end{cases}
\ee
Mapping this solution back to the orbifold branch, generic Dirichlet boundary conditions are grouped into orbits \cite{Oshikawa:1996dj}:
\be
\big|D^+, \theta \big\rangle = \frac{|D, \theta \big\rangle + \big|D, -\theta \big\rangle}{\sqrt{2}} \, ,
\ee
and have $g$-function equal to $\sqrt{1/R_{\text{orb}}}$. The IR defect becomes:
\be
\calD_{IR} = 
   \Big|D^+, \frac{3\pi}{4}\Big\rangle \Big\langle D^+, \frac{3 \pi}{4}\Big| \, , \ \ \ \lambda > 0  \ \ \ \ \ \ \ 
   \calD_{IR} = \Big|D^+, \frac{\pi}{4}\Big\rangle \Big\langle D^+, \frac{\pi}{4}\Big| \, , \ \ \ \lambda < 0 \, ,
\ee
with $g$-function $
g\left(\calD_{IR}\right) = \frac{1}{R_{\text{orb}}}
$, consistently with the $g$-theorem.
This is yet again a factorized defect.

\section{Surface defects in $(2+1)$d scalar QED}\label{sec: 3+1}

The aim of this section is to study the IR dynamics of the defects we introduced in \ref{sec:QED3} in certain limits ($m^2\rightarrow \infty$ and $N_f\gg 1$) to explicitly verify the prediction from the symmetry reflecting nature of the defect. The special feature of these limits is that the bulk theory is quadratic in the photon: it is 3d Maxwell (hence free) for $m^2\rightarrow \infty$\footnote{In this case of free bulk, our surface defect has been recently analyzed in \cite{Fraser-Taliente:2024lea}, showing that it flows to a nontrivial DCFT in the infrared.}, while for large $N_f$ limit, at leading order in $N_f$ the bulk is a generalized free theory in which the photon propagator is obtained by resumming bulk scalar bubbles. For large $N_f$ we tune the scalar mass to its critical value, hence the bulk flows to a CFT where the photon has scaling dimension $1$. 
By our general arguments on symmetry reflecting defects, we still expect the defect to be non-decoupled at the end of the defect RG flow. 

In both cases the general strategy is to integrate out the bulk photon to obtain the exact effective action on the defect that, in momentum space, takes the general form  (see Appendix \ref{app:conti} for the detailed computation)
\begin{equation}
    S_{\text{def}, \text{eff}} = \frac{1}{4\pi}\int \frac{d^2 p}{(2\pi)^2}\, p^2 R^2(p^2) \ \phi(p)\phi(-p)\, ,
\end{equation}
where the expression \eqref{eq: effectiveradius} of the effective radius $R^2(p)$ is determined in Appendix \ref{app:conti}, and is different in the two cases.
In position space the defect theory generically contains non-local (long-range) terms, induced by the $3d$ photon, that can influence the IR behavior. We also compute correlation functions of bulk and defect operators, in particular the two point function $\big\langle \partial_i \phi(x)\,  F_{\mu \nu}(0) \big\rangle$ that allows us to establish the predicted non-decoupling at low energy.

\subsection{Surface defect in $3d$ Maxwell theory}\label{ssec: maxwell}
The propagator of the free Maxwell field in the Coulomb gauge takes the form 
\begin{equation}\label{eq:photonprop}
    G_{\mu \nu}(p) = \left(\delta_{\mu \nu}-\frac{p_\mu p_\nu}{p^2}\right)\Pi(p^2)\, , \ \ \ \ \ \ \ \   \Pi(p^2) = \frac{e^2}{p^2}\, .
\end{equation}
This leads to the effective radius (see Appendix \ref{app:conti} for the derivation):
\begin{equation}
    R^2(p^2) = R^2 + \frac{e^2}{4 \pi |p|}\, .
\end{equation}
In the UV, $|p|\gg e^2$, $R^2(p^2)$ reduces to the original radius $R$, while in the IR limit the second term dominates. This different infrared behavior implies that the $2d$ scalar $\phi$ acquires an anomalous dimension:
\begin{equation}
    \gamma_\phi =\frac{1}{2} \ .
\end{equation}
For $|x-y|e^2\gg 1$ we have (see \cite{Fraser-Taliente:2024lea} for the exact expression of the propagator)
\begin{equation}
   \langle \phi(x)\phi(y)\rangle \sim \frac{1}{|x-y|}\, .
\end{equation}
Similarly, we can compute the correlator of gauge invariant operators:
\begin{equation}\label{eq: 3d Maxwell 2pt phi}
     \big\langle \, \partial_i\phi(x)\, \partial_j\phi(y)\, \big\rangle \sim \left(\delta_{ij} - 3\frac{(x_i-y_i)( x_j-y_j)}{|x-y|^2}\right)\frac{1}{|x-y|^3}\, .
\end{equation}
Therefore the defect QFT flows to a nontrivial DCFT in the IR, consistently with the UV defect being symmetry reflecting. 

Using the generating functional computed in App. \ref{app:conti} we can access the two-point function of the current $\partial_i \phi$ and the bulk field strength $F_{\mu \nu}$. We get, in the IR limit on the defect: 
\begin{equation} \label{eq: OPE 3d Maxwell}
  i\big\langle \, p_k \phi(p)\,  F_{i \perp}(-p, z)\, \big\rangle = -2 \pi \epsilon_{ij} \frac{p^j p_k}{|p|} \exp\left(-|p| |z|\right) \sign(z) \simeq -2 \pi\frac{\epsilon_{ij}p^jp_k}{|p|} + O(z) \, .
\end{equation}
In position space instead we find: 
\begin{equation}
    \langle \partial_k \phi(x) F_{i \perp}(0, 0)\rangle = -\epsilon_{ij}\left(\delta_{kj} - 3\frac{x_k x_j}{|x|^2}\right) \frac{z^2 -|x|^2}{(z^2 +|x|^2)^{5/2}} \sign(z) \sim \epsilon_{ij}\left(\delta_{kj} - 3\frac{x_k x_j}{|x|^2}\right)\frac{1}{|x|^3} + O(z)\, .
\end{equation}
By introducing the dual scalar $\varphi$, defined as
\begin{equation}
    \partial_i \varphi := \frac{1}{e}\epsilon^{ij}F_{j\perp}\,,
\end{equation}
we can rewrite the correlator \eqref{eq: OPE 3d Maxwell} as the two point function $\langle \partial_k \phi(x) \partial_j\varphi (0, 0)\rangle$. By comparing this result with \eqref{eq: 3d Maxwell 2pt phi}, we thus conclude that the leading order contribution to the bulk-defect OPE of the dual scalar $\varphi$ is given by the compact scalar $\phi$ on the defect. As the IR dimension of $\phi$ matches that of a free $3d$ scalar there are no divergencies in the bulk-defect OPE. This confirms that the defect we are studying flows to a gapless interacting defect in the IR.

\subsection{Surface defect in large $N_f$ QED$_3$}

The large $N_f$ limit of scalar QED$_3$, at leading order, is a free theory in which, with an appropriate choice of gauge, the photon propagator takes the form
\begin{equation}\label{eq:photonpropNF}
 G_{\mu \nu}(p) = \left(\delta_{\mu \nu}-\frac{p_\mu p_\nu}{p^2}\right)\Pi(p^2)\, , \ \ \ \ \ \ \ \     \Pi(p^2) = \frac{16 \alpha}{N_f}\frac{1}{p^2 + \alpha |p|} \ .
\end{equation}
Here $\alpha \sim e^2 N_f$ is the 't Hooft coupling. By tuning the scalar masses, in the IR the bulk theory is a CFT in which the photon has dimension $1$ with $\Pi(p^2)\sim 1/|p|$. 

We want to study the surface defect of Section \ref{sec:QED3} in this theory. There are obviously two different RG flows: we can either couple the 2d compact boson to the UV limit of scalar QED (fine tuned to flow to criticality) and make the defect and the bulk to flow together, or we can  directly couple the defect in the critical bulk theory, hence studying the defect RG flow. We will see shortly that the IR theories coincide, hence we can consider either one to test the non-decoupling of the defect. 
To study the first type of flow we consider \eqref{eq: effectiveradius} using the full propagator \eqref{eq:photonpropNF}:
\begin{equation}\label{eq:effRUV}
\begin{split}
      R^2(p^2) &= R^2 + \frac{4 \alpha}{\pi^2 N_f}\int_{-\infty}^{\infty} dp_{\perp} \frac{1}{p^2 + p_{\perp}^2 + \alpha \sqrt{p^2 + p_{\perp}^2 }} \\ & = R^{2} - \frac{8}{\pi^2 N_f}\frac{\alpha}{|p|} \frac{\log\left(\alpha/ |p|-\sqrt{\left(\alpha /|p|\right)^2-1}\right)}{\sqrt{\left(\alpha /|p|\right)^2-1}}\, ,
\end{split}
\end{equation}
which is real and positive for all positive values of $|p|$. The effective radius interpolates between the two regimes
\begin{equation}\label{eq:RUVIR}
    \begin{array}{ll}
      \displaystyle   R^2_{\text{UV}}(p)= R^2 + \frac{4}{\pi N_f}\frac{\alpha}{|p|}+ O\left(\frac{\alpha^2}{p^2}\right)\, , & \ \ \ \ \ |p|\gg \alpha \vspace{0.4cm} \\ 
      \displaystyle   R^2_{\text{IR}}(p)= R^2 + \frac{8}{\pi^2 N_f}\log\left(\frac{2\alpha}{|p|}\right)+ O\left(\frac{p^2}{\alpha^2}\log\left(\frac{|p|}{\alpha}\right)\right)\, , & \ \ \ \ \  |p|\ll \alpha\, .
    \end{array}
\end{equation}
In the UV we recover the compact boson at radius $R$, while in the IR the effective radius diverges logarithmically. 
In the second UV definition we use \eqref{eq: effectiveradius} with the infrared photon propagator: 
\begin{equation}
    R^2(p^2) = R_0^2 + \frac{8}{\pi N_f}\int_{-\infty}^{\infty} \frac{dp_\perp}{(2\pi)} \frac{1}{\sqrt{p^2 + p_\perp^2}} \, .
\end{equation}
$R_0$ is the bare radius. The integral over the transverse momentum is UV divergent and requires regularization. A neat way to achieve this is to use dimensional regularization for the defect's world-volume dimension $2 + \epsilon$ \cite{deSabbata:2024xwn}:
\begin{equation}\label{eq:effRdiv}
\begin{split}
  R^2(p^2) &= R_0^2 + \frac{8}{\pi N_f}\frac{\Gamma\left(\frac{\epsilon}{2}\right) \Gamma\left(\frac{1-\epsilon}{2}\right)}{2\sqrt{\pi}(2\pi)^{(1-\epsilon)}}\left(\frac{\mu}{|p|}\right)^{\epsilon} \\ & = R^2 + \frac{4}{\pi^2 N_f}\log\left(\frac{\mu}{|p|}\right) = \frac{4}{\pi^2 N_f}\log\left(\frac{\Lambda}{|p|}\right)\, ,
\end{split}
\end{equation}
so that the divergence can be reabsorbed in the definition of the bare radius.\footnote{This is equivalent to cancel the UV divergence of the integral fixing a renormalization condition $R^2(|p|=\mu) = R^2$. More precisely we can set $R_0 = Z_R R \mu ^{\frac{d-1}{2}}$ and $\phi_0 = Z_{\phi} \phi \mu^{\frac{1-d}{2}}$, then we can reabsorb the divergence requiring that the propagator at  $|p|=\mu$ coincides with that of a compact boson of radius $R$. We can then choose to work in a scheme in which $Z_\phi=1$ and hence $\phi=\phi_0$. }
In the last equality we have introduced a UV scale $\Lambda=\mu e^{\frac{\pi^2 N_f}{4}R^2}$ at which the effective radius vanishes. This new scale behaves much like the standard Landau pole in $4$ dimensional $\phi^4$ theory: it is a new dimensional parameter of the quantum theory that arises in place of the classically dimensionless radius and sets a cutoff below which the theory is well-defined.  Indeed here $R^2(p^2)$ is positive only for $|p|<\Lambda$. A defect with similar behavior also arises in $4d$ Maxwell theory coupled to a $2d$ compact boson \cite{Fraser-Taliente:2024lea}. 

Imposing independence on $\mu$ of the effective radius gives a constant beta function
\begin{equation}
    \mu\frac{d R^2}{d \mu } = - \frac{8}{\pi^2 N_f}\, ,
\end{equation}
it is easy to see, computing the beta function for the combination $\lambda_{\text{PBN}}= R^2/(1+R^2)$, that the only fixed point is at $R^2=\infty$ and is reached in the deep IR.\footnote{The new coupling $\lambda_{\text{PBN}} \in [0,1]$ is particularly useful for analyzing fixed points in the context of positive, real-valued couplings. Here its beta function is $-\dfrac{8}{\pi^2 N_f}(1-\lambda_{\text{PBN}})^2$ that has a unique attractive fixed point at $\lambda_{\text{PBN}}=1$.}
We expect that such a state of affairs is a consequence of the free nature of the bulk theory for large $N_f$, and the fixed point  might move at finite values of $R^2$ once $1/N_f$ corrections are taken into account. It would be interesting to explore this aspect in the future.

Comparing the two definitions of the defect we see that the theory with effective radius \eqref{eq:effRUV} can be thought of as a UV completion of the defect QFT defined by \eqref{eq:effRdiv} in which the Landau pole scale $\Lambda$ is fixed by the dimensional bulk coupling $\alpha$ and the UV radius $R$.
The IR behavior of the effective radius, however, is the same whether we use \eqref{eq:effRUV} or \eqref{eq:effRdiv}, hence they define the same IR DQFT at scales $|p|\ll \Lambda$. 

Let us further analyze the IR theory looking at two point functions in position space. We will use the same regularization scheme and take the limit $\epsilon \to 0$ at the end of the computation. Furthermore, in order to avoid further divergences it is convenient to perform wavefunction renormalization on $\phi$ by mapping it to $R_0 \phi$.
We consider the correlator
\begin{equation}
    \langle \phi(x)\phi(y) \rangle =\int\frac{d^{2+\epsilon} p}{(2\pi)^{2+\epsilon}} \frac{e^{ip\cdot(x-y)}}{p^2} \frac{1}{1 +\frac{8}{\pi  N_f} \frac{\kappa_\epsilon}{R_0^2} \left(\frac{\mu}{|p|}\right)^{\epsilon} }
\end{equation}
where $\kappa_{\epsilon} =\frac{\Gamma\left(\frac{\epsilon}{2}\right) \Gamma\left(\frac{1-\epsilon}{2}\right)}{2\sqrt{\pi}(2\pi)^{(1-\epsilon)}}$.
At large distances $|x-y|\mu \gg 1 $ and for $\epsilon>0$, the second term in the denominator dominates and we find
\begin{equation}
   \langle \phi(x)\phi(y) \rangle \simeq \frac{\pi N_f}{8}\frac{R_0^2}{\kappa_d}\mu^{-2 \epsilon}\int\frac{d^{2+\epsilon} p}{(2\pi)^{2+\epsilon}} e^{ip\cdot(x-y)}|p|^{-2 +\epsilon}\, ,
\end{equation}
which gives:\footnote{We are absorbing an infinite additive constant in the definition of the scale $\mu$. Such a divergence also appears in the naive free theory analysis in Fourier space, and at any rate vanishes in the correlation functions of well defined operators, such as currents.}
\begin{equation}
    \langle \phi(x)\phi(y) \rangle = \frac{1}{2}\log\left(\mu |x-y|\right)\, \, .
\end{equation}  
We have used that $\frac{8}{\pi N_f}\kappa_\epsilon/R_0^2 \sim -1 + O(\epsilon)$ and absorbed finite scheme dependent terms in the definition of $\mu$. 
This implies that $\phi$ has vanishing dimension at the IR fixed point, compatible with a free scalar field. 

An alternative way to obtain this result is to write the $2$-point function of $\phi$ as the integral
\begin{equation}
    \langle \phi(x)\phi(y) \rangle =\int_0^\infty \frac{d p}{(2\pi)} \frac{J_0(p|x|)}{p} \frac{1}{R^{2}(p^2)}\, , 
\end{equation}
where $J_0$ is the zeroth order Bessel function and $R^2(p^2)$ is the effective radius obtained either using the UV completion or an hard cutoff regularization.\footnote{In this case
\begin{equation}\label{eq:HCR}
    R^{2}(p^2) = R^2 + \frac{4}{\pi^2 N_f}\log\left(1 + \frac{2 \Lambda\left(\Lambda+ \sqrt{p^2 + \Lambda^2}\right)}{p^2}\right)
\end{equation}
which is positive for all $|p|\ge 0$ and has the proper logarithmic behavior for $|p|\ll \Lambda$. } We will not employ the dimensionally regularized effective radius as it always vanishes at some UV scale and we want to avoid poles along the integration contour. The integral involving $J_0$ is formally IR divergent due to the inverse power of $p$, we can regulate it taking a derivative in $|x|$ and using the identity
\begin{equation}
    \frac{d}{d|x|}J_0(p |x|) = - p J_1(p |x|)\, .
\end{equation}
We can then study the integral
\begin{equation}\label{eq:intJ1}
    \int_0^\infty \frac{d p}{(2\pi)}  \frac{J_1(p|x|)}{R^{2}(p^2)}\, , 
\end{equation}
which is convergent in both UV and IR and can be related to the two-point function involving $\phi$ and its first derivative. Picking an explicit form of $R^2(p^2)$, either the UV completed one \eqref{eq:effRUV} or the hard cutoff regulated one \eqref{eq:HCR}, the integral \eqref{eq:intJ1} can be computed numerically. It is possible to check explicitly that, for large $|x|\Lambda$, where $\Lambda$ is the dimensionful scale appearing in $R^2(p^2)$, the integral indeed decreases as $1/|x|$, which is compatible with the logarithmic correlator above.

As in the previous subsection the simplest observable that probes the bulk-defect interaction is the two point function involving the bulk field strength and the current $\partial_i\phi$. Starting from the momentum space expression \eqref{eq:2ptpF} and Fourier transforming in $p_{\perp}$ we get 
\begin{equation}
  i\langle p_k \phi(p) F_{i \perp}(-p, z)\rangle \sim \frac{\epsilon_{ij}p_k p^j}{p^2 R^2(p^2)}|p|K_1(|p| |z|) \sign(z) \sim\frac{1}{z}\frac{\epsilon_{ij}p_k p^j}{p^2 R^2(p^2)}\, .
 \end{equation}
In the last equality we expanded the modified Bessel function $K_1(|p|z)$ for small values of $z$ and only kept the leading term. Passing in position space for the remaining directions we can use our previous results to conclude that 
\begin{equation}
    \langle \partial_k \phi(x) F_{i \perp}(0, z)\rangle 
    \sim \frac{1}{z}\partial_k \epsilon_{ij}\partial_j \log\left(\mu |x|\right)\sim \frac{\epsilon_{ij}}{x^2}\left(\delta_{kj} -2\frac{x_k x_j}{x^2}\right)\frac{1}{z} + ... \, ,
\end{equation}
where the dots denote finite terms in the small $z$ limit.
This is consistent with the bulk-defect OPE 
\begin{equation}
    F_{i \perp} \sim \frac{1}{z} {j_{W}}_i + ... \, ,
\end{equation}
with $\star j_W = \frac{d \phi}{2\pi}$ the winding current.
It would be interesting to study the defect OPE of monopole operators. Following our discussion in Section \ref{sec:QED3}, it is clear that the bulk $U(1)_T$ must be matched by the defect's momentum symmetry. In the UV description this is carried by vertex operators $V_n =e^{i n \phi}$, whose fate, however is far from clear in the low energy description. Another interesting aspect concerns the physics of dressed vortices on the defect's world-volume. Clearly, fractionalized vortices cannot appear in any OPE with bulk field, whose transformation under the center of $SU(N_f)$ is always trivial. 
We leave these problems for future work.

\paragraph{Acknowledgments} We thank Mohamed Anber, Matthew Bullimore, Christopher Herzog, Ho-Tat Lam, Mark Mezei, Pierluigi Niro, Diego Rodriguez-Gomez, Marco Serone, Ritam Sinha, Tin Sulejmanpasic, Luigi Tizzano and Yifan Wang for discussions. 
The work of  A.A. is supported by the UKRI Frontier Research Grant, underwriting the ERC Advanced Grant ''Generalized Symmetries in Quantum Field Theory and Quantum Gravity''. C.C. is supported by STFC grant ST/X000761/1. The research of G.G. is funded through an ARC advanced project, and further supported by IISN-
Belgium (convention 4.4503.15).
We thank NYU and the Simons Collaboration for Categorical Symmetries for hospitality during the annual meeting, where some of this work was finalized.

\appendix

\section{$\bZ_2$ twisted sectors of unitary minimal models}\label{app: Z2 minimal models}

In this appendix we compute the $\bZ_2$ charge of the states living in the $\bZ_2$-twisted Hilbert space of the diagonal invariant minimal models. In particular we want to find $\bZ_2$ invariant operators in the twisted sector of the $\bZ_2$ symmetry itself. For the sake of brevity, we only report the relevant computations and we assume the standard CFT language (see e.g. \cite{DiFrancesco:1997nk}).
The untwisted partition function of the models is
\be
\cZ_m = \sum_{(r,s)}\chi_{r,s}(\tau) \chi_{r, s}(\overline{\tau})\,
\ee
where $r,s$ are restricted to the range $1\leq r< m$, $1\leq s < m+1$ with the identification $(r,s) \sim (m+1-r, m-s)$.
The $\bZ_2$ twisted partition function, twisted by the Verlinde line $L_{1,m}$ is
\begin{equation}
    \cZ_2 (\tau) = \sum_{(r,s)}\chi_{r,m-s+1}(\tau) \chi_{r, s}(\overline{\tau})\, 
\end{equation}
from which we extract the dimensions and spins of the $\bZ_2$-twisted sector operators as
\begin{equation}
    \begin{split}
        &\Delta_{r,s} = h_{r,m-s+1}+ h_{r,s} = 2 h_{r,s} + \frac{(m-2r)(m-2s+1)}{4}\\
        & J_{r,s} =  h_{r,m-s+1}- h_{r,s} = \frac{(m-2r)(m-2s+1)}{4}\in \frac{1}{2}\bZ\,.
    \end{split}
\end{equation}
Using a $T$ transformation, we can map the twisted partition function to the twisted partition function, twined by a $\bZ_2$ line. From this expression, we see that the $\bZ_2$ invariant operators are only those with integer spins. The constraints that this imposes on $r,s$ depend on $m \bmod 4$, in particular
\begin{itemize}
    \item $m = 0 \bmod 4$. We have
    \begin{equation}
        J_{r,s} = \frac{r}{2} \bmod 1\,.
    \end{equation}
    Therefore all operators with $r$ even are $\bZ_2$ singlets, the first corresponds to $(r,s)= (2,1)$ has
    \begin{equation}
    \begin{split}
        &\Delta_{2,1}= \frac{6 + m (m-2)(m-3)}{4 m}\\
        & J_{2,1} = \frac{(m-4)(m-1)}{4}\, .
    \end{split}
    \end{equation}
    The lightest operator has $(r,s)=(m/2, m/2)$ with
    \begin{equation}
        \Delta_{m/2, m/2} = \frac{m^2-4}{8m(m+1)}
    \end{equation}
    and vanishing spin. The first few values are
    \begin{equation}
        \Delta_{m/2, m/2}= \frac{3}{40}, \frac{5}{48}, \frac{35}{312}
    \end{equation}
    for $m=4, 8, 12$ respectively. These operators are always relevant in $1$ dimension. 
    
\item $m = 1 \bmod 4$. We have
    \begin{equation}
        J_{r,s} = \frac{1-s}{2} \bmod 1\,.
    \end{equation}
    Therefore all operators with $s$ odd are $\bZ_2$ singlets, the first corresponds to $(r,s)= (1,1)$, which is chiral and has
    \begin{equation}
    \begin{split}
        &\Delta= J = \frac{(m-2)(m-1)}{4}\, .
    \end{split}
\end{equation} 
The lightest operator has $(r,s)=((m+1)/2, (m+1)/2)$ with
    \begin{equation}
        \Delta_{(m+1)/2, (m+1)/2} = \frac{(m+3)(m-1)}{8m(m+1)}
    \end{equation}
    and vanishing spin. The first few values are
    \begin{equation}
       \Delta_{(m+1)/2, (m+1)/2}= \frac{2}{15}, \frac{2}{15}, \frac{12}{91}
    \end{equation}
    for $m=5, 9, 13$ respectively. These operators are always relevant in $1$ dimension. 

\item $m = 2 \bmod 4$. We have
    \begin{equation}
        J_{r,s} = \frac{1-r}{2} \bmod 1\,.
    \end{equation}
    Now all operators with $r$ odd are $\bZ_2$ singlets. The first is again $(r,s)= (1,1)$, which is chiral and has
    \begin{equation}
    \begin{split}
        &\Delta= J = \frac{(m-2)(m-1)}{4}\, .
    \end{split}
\end{equation}
The lightest operator is $(r,s)=(m/2, m/2)$ 
    \begin{equation}
        \Delta_{m/2, m/2}= \frac{m^2-4}{8m(m+1)}
    \end{equation}
    and vanishing spin. The first few values are
    \begin{equation}
       \Delta_{m/2, m/2}= \frac{2}{21}, \frac{6}{55}, \frac{4}{35}
    \end{equation}
    for $m=6, 10, 14$ respectively. These operators are always relevant in $1$ dimension. 

\item $m = 3 \bmod 4$. We have
    \begin{equation}
        J_{r,s} = \frac{s}{2} \bmod 1\,.
    \end{equation}
    now all operators with $s$ even are $\bZ_2$ singlets. The first is $(r,s)= (1,2)$ which has
    \begin{equation}
    \begin{split}
        &\Delta= \frac{2 + m (m-1)(m-3)}{4 (m+1)}\\
        & J = \frac{(m-2)(m-3)}{4}\, .
    \end{split}
\end{equation}
 The lightest operator is $(r,s)= ((m+1)/2,(m+1)/2) $ with
 \begin{equation}
     \Delta_{(m+1)/2,(m+1)/2} = \frac{(m-1)(m+3)}{8m(m+1)}
 \end{equation}
and vanishing spin. The Ising CFT corresponds to $m=3$ and the numbers above specify to $\Delta = 1/8$ and $J=0$ correctly for the $\mu$ operator used in the main text.
\end{itemize}
Therefore for any $m$ we find a relevant (in 1 dimension) scalar operator labelled by
\begin{equation}
    (r,s) = \left(\left\lceil\frac{m}{2}\right \rceil, \left\lceil\frac{m}{2}\right \rceil\right)
\end{equation}
and
\begin{equation}
    \Delta_{\left\lceil\frac{m}{2}\right \rceil, \left\lceil\frac{m}{2}\right \rceil} =\begin{cases}&\frac{m^2-4}{8m(m+1)} \text{ if } m =0 \bmod 2 \\ & \frac{(m-1)(m+3)}{8m(m+1)} \text{ if } m =1 \bmod 2\end{cases} = \frac{1}{8},\frac{3}{40},\frac{2}{15},\frac{2}{21},\frac{15}{112},...
\end{equation}
for $m=3,4,5,6,7,..$. Therefore, there always is a relevant $\mathbb{Z}_2$ invariant operator in the $\bZ_2$ twisted sector. 

\section{Generating functional for DQFT correlators}    \label{app:conti}
In this Appendix we collect some of the technical material needed to study defect dynamics. 

One of the general situation we consider in the main text can be described by the following action
\begin{equation}
    S = S_{\text{bulk}} + \int_{x_\perp=0} d^{d-1}x\,  A_{\mu}J^{\mu} + S_{\text{defect}}
\end{equation}
where we assume that the bulk theory contains a photon $A_{\mu}$ that couples to the $d-1$ dimensional theory via a current $J^{\mu}$. The bulk theory is generally interacting, here we assume that the effects of these interactions can, in some limit, be represented via an effective photon propagator $G_{\mu \nu}$. Then we effectively integrate out the photon exactly and write
\begin{equation}
    S= -\frac{1}{2}\int d^d x d^d y \left(K^\mu + J^\mu\right) G_{\mu \nu}(x-y)\left(K^\nu + J^\nu\right) + S_{\text{def}}
\end{equation}
where we have added a source $K^{\mu}$ for the photon. On general grounds we would need to compute the generating functional for the photon correlation functions with a fixed source turned on, corresponding to the defect current. As long as the photon behaves as a free field, namely we can compute correlators using the Wick theorem, the computation is straightforward and the result is the above. Let us denote by $\sigma_i$ the coordinates parallel to the defect, then we have $J^\perp=0$ and $J^i= \widetilde{J}^i(\sigma) \delta(x^d)$. The action is then
\begin{equation}
\begin{split}
     S  =& -\frac{1}{2}\int d^d x d^d y K^\mu(x) G_{\mu \nu}(x-y)K^\nu(y)  -\int d^d x d^{d-1}\sigma K^\mu(x) G_{\mu i}(x-\sigma)\widetilde{J}^i(\sigma)  \\ & -\frac{1}{2}\int d^{d-1} \sigma d^{d-1}\sigma' \widetilde{J}^i(\sigma) G_{i j}(\sigma-\sigma')\widetilde{J}^j(\sigma')+ S_{\text{def}}\, .
\end{split}
\end{equation}
For the theories considered in Section \ref{sec: 3+1} we need to fix $d=3$ and 
\begin{equation}
\begin{split}
    &S_2 = \frac{R^2}{4\pi}\int d^2\sigma \partial_i \phi \partial^i \phi\\
    & \widetilde{J}_i(\sigma) = \frac{i}{2\pi}\epsilon_{ij}\partial^j \phi\, .
\end{split}
\end{equation}
In momentum space we have 
\begin{equation}
\begin{split}
    S=&  -\frac{1}{2} \int \frac{d^3 p}{(2\pi)^3} K_{\mu}(p) G^{\mu \nu}(p)K_{\nu}(p)+ \frac{1}{2\pi} \int \frac{d^2 p dx}{(2\pi)^3} K^{\mu}(-p,-x)G_{\mu i}(p,x) \epsilon^{ij}p_j \phi(p)\\ & + \frac{1}{8\pi^2}\int \frac{d^2 p}{(2\pi)^2} \epsilon_{ij}p^j \left[\int \frac{dx}{2\pi} G^{ik}(p, x)\right]\epsilon_{kl}p^l \phi(-p)\phi(p) + \frac{R^2}{4\pi}\int \frac{d^2 p}{(2\pi)^2}p^2 \phi(p)\phi(-p)\, .
\end{split}
\end{equation}
Let us also turn on a source $Y(p)$ for the scalar field $\phi(p)$\footnote{In order for the coupling to be well defined we require that $Y(0) \in \bZ$, which is equivalent to requiring the function in position space to have quantized integral.} so that 
\begin{equation}
\begin{split}
    S &= -\frac{1}{2} \int \frac{d^3 p}{(2\pi)^3} K_{\mu}(p) G^{\mu \nu}(p)K_{\nu}(p)  \\ & -\int \frac{d^2 p}{(2\pi)^2}\left[Y(-p)-\frac{1}{2\pi}\int \frac{d x}{2\pi}K^{\mu}(-p,-x)G_{\mu i}(p,x) \epsilon^{ij}p_j \right] \phi(p)\\ &+ \int \frac{d^2 p}{(2\pi)^2} \frac{R^2 p^2}{4\pi}\left[1+ \frac{1}{2\pi}\frac{\epsilon_{ij}p^j\epsilon_{kl}p^l}{R p^2} \left[\int \frac{dx}{2\pi} G^{ik}(p, x)\right]\right]\phi(p)\phi(-p)\, .
\end{split}
\end{equation}
Finally, integrating out $\phi(p)$, we obtain the generating functional
\begin{equation}
\begin{split}
    W[K, Y] =&  -\frac{1}{2} \int \frac{d^3 p}{(2\pi)^3} K_{\mu}(p) G^{\mu \nu}(p)K_{\nu}(p) - \frac{1}{2} \int \frac{d^2p}{(2\pi)^2}\frac{2\pi}{p^2 R^2(p^2)}Y(-p)Y(p) \\ & +\frac{1}{2\pi}\int \frac{d^2 p dx}{(2\pi)^3}\frac{2\pi}{p^2 R^2(p^2)}Y(-p)K^\mu(p,x)G_{\mu i}(p,x)\epsilon^{ij}p_j\\ & -\frac{1}{8\pi^2}\int \frac{d^2 p dx dy} {(2\pi)^5}K^\mu(p,x)G_{\mu i}(p,x)\epsilon^{ij}p_jK^\nu(p,y)G_{\nu k}(p,y)\epsilon^{kl}p_l\, ,
\end{split}
\end{equation}
where
\begin{equation}
R^2(p^2) = R^2 + \frac{1}{2\pi}\frac{\epsilon_{ij}p^j\epsilon_{kl}p^l}{p^2} \int \frac{dp_{\perp}}{2\pi} G^{ik}(p, p_{\perp})\, .
\end{equation}
Taking functional derivatives we can compute connected correlators of the DQFT. To simplify the expressions we are going to assume a choice of gauge such that the photon propagator is transverse
\begin{equation}\label{eq:photonpropapp}
    G_{\mu \nu}(p) = \left(\delta_{\mu \nu}-\frac{p_\mu p_\nu}{p^2}\right)\Pi(p^2)\, , 
\end{equation}
so that
\begin{equation} \label{eq: effectiveradius}
R^2(p^2) = R^2 + \frac{1}{2\pi}\int_{-\infty}^{+\infty} \frac{dp_\perp}{2\pi} \Pi(p^2+ p_\perp^2)\, .
\end{equation}
Diagrammatically the effective radius is obtained resumming photon ``half-bubbles" in the defect propagator, schematically
\be
\begin{tikzpicture}[baseline={(0,0.5)}]
\draw[dashed] (0,0) -- (2,0) -- (3,1) -- (1,1) --cycle;   
\draw[line width=2, blue] (0.5,0.5) -- (2.5,0.5);
\end{tikzpicture}
=
\begin{tikzpicture}[baseline={(0,0.5)}]
 \draw[dashed] (0,0) -- (2,0) -- (3,1) -- (1,1) --cycle;   
\draw (0.5,0.5) -- (2.5,0.5);
\end{tikzpicture}
+
\begin{tikzpicture}[baseline={(0,0.5)}]
 \draw[dashed] (0,0) -- (2,0) -- (3,1) -- (1,1) --cycle;   
\draw (0.5,0.5) -- (1,0.5); \draw (2,0.5) -- (2.5,0.5);
\draw[snake it] (1,0.5) arc (180:0:0.5 and 0.75);
\draw[fill=black] (1,0.5) circle (0.05);
\draw[fill=black] (2,0.5) circle (0.05);
\end{tikzpicture}
+
\begin{tikzpicture}[baseline={(0,0.5)}]
 \draw[dashed] (0,0) -- (2,0) -- (3,1) -- (1,1) --cycle;   
\draw (0.5,0.5) -- (0.75,0.5); \draw (2.25,0.5) -- (2.5,0.5); \draw (1.25,0.5) -- (1.75,0.5);
\draw[snake it] (0.75,0.5) arc (180:0:0.25 and 0.75); \draw[snake it] (1.75,0.5) arc (180:0:0.25 and 0.75);
\draw[fill=black] (0.75,0.5) circle (0.05);
\draw[fill=black] (1.25,0.5) circle (0.05);
\draw[fill=black] (1.75,0.5) circle (0.05);
\draw[fill=black] (2.25,0.5) circle (0.05);
\end{tikzpicture}
+ . . . 
\ee
where the wiggly lines denote the bulk photon propagator. Notice that the diagrams are tree level only and the integral comes from the momentum perpendicular to the defect. The $Y-K$ mixed term in the generating functional corresponds to the two point function $ \langle \phi(p) A_{\mu}(-p, p_\perp)\rangle $ from which we can extract the gauge invariant correlator
\begin{equation}\label{eq:2ptpF}
  \langle p_k \phi(p) F_{i \perp}(-p, p_\perp)\rangle = p_{\perp}\epsilon_{ij}p^jp_k\frac{\Pi\left(p^2 + p_{\perp}^2\right)}{p^2 R^2(p^2)}\, .
\end{equation}
Passing in position space for the perpendicular direction we can study the bulk-defect OPE of the field strength $F_{\mu \nu}$, which probes the IR bulk-defect coupling.

\small{
\bibliographystyle{ytphys}
\baselineskip=0.85\baselineskip
\bibliography{bibgw}

}
\end{document}